\documentclass[aps,pra,showpacs,amsmath,amssymb,reprint,nofootinbib]{revtex4-1}

\usepackage{graphicx}
\usepackage{dcolumn}
 \AtBeginDocument{\usepackage{booktabs}}
\usepackage{units}
\usepackage{hyperref}

\newcommand{\ionm}[2]{${}^{#1}$#2${}^+$}
\newcommand{\ion}[1]{#1${}^+$}

\begin{document}

\preprint{}

\title[]{Controlling systematic frequency uncertainties at the $10^{-19}$ level in linear Coulomb crystals}

%PACS: 37.10.Ty is Atomic physics: Ion trapping, 06.30.Ft is Metrology: Time and frequency

\author{J.~Keller}
\affiliation{Physikalisch-Technische Bundesanstalt, Bundesallee 100, 38116 Braunschweig, Germany}
\author{T.~Burgermeister}
\affiliation{Physikalisch-Technische Bundesanstalt, Bundesallee 100, 38116 Braunschweig, Germany}
\author{D.~Kalincev}
\affiliation{Physikalisch-Technische Bundesanstalt, Bundesallee 100, 38116 Braunschweig, Germany}
\author{A.~Didier}
\affiliation{Physikalisch-Technische Bundesanstalt, Bundesallee 100, 38116 Braunschweig, Germany}
\author{A.~P.~Kulosa}
\affiliation{Physikalisch-Technische Bundesanstalt, Bundesallee 100, 38116 Braunschweig, Germany}
\author{T.~Nordmann}
\affiliation{Physikalisch-Technische Bundesanstalt, Bundesallee 100, 38116 Braunschweig, Germany}
\author{J.~Kiethe}
\affiliation{Physikalisch-Technische Bundesanstalt, Bundesallee 100, 38116 Braunschweig, Germany}
\author{T.~E.~Mehlst\"aubler}
\affiliation{Physikalisch-Technische Bundesanstalt, Bundesallee 100, 38116 Braunschweig, Germany}

\date{\today}

\begin{abstract}
Trapped ions are ideally suited for precision spectroscopy, as is evident from the remarkably low systematic uncertainties of single-ion clocks. The major weakness of these clocks is the long averaging time, necessitated by the low signal of a single atom. An increased number of ions can overcome this limitation and allow for the implementation of novel clock schemes. However, this presents the challenge to maintain the excellent control over systematic shifts of a single particle in spatially extended and strongly coupled many-body systems. We measure and deduce systematic frequency uncertainties related to spectroscopy with ion chains in a newly developed rf trap array designed for precision spectroscopy on simultaneously trapped ion ensembles. For the example of an \ion{In} clock, sympathetically cooled with \ion{Yb} ions, we show in our system that the expected systematic frequency uncertainties related to multi-ion operation can be below $1\times10^{-19}$. Our results pave the way to advanced spectroscopy schemes such as entangled clock spectroscopy and cascaded clock operation.
\end{abstract}

\maketitle
\section{Introduction}
State-of-the art ion optical clocks have demonstrated fractional systematic uncertainties approaching $1\times10^{-18}$ ~\cite{Chou2010a,Dube2014,Huntemann2016}, thus being among the most accurate artificially made devices. Ion traps provide strong confinement for charged particles without influencing their electronic energy levels to first order. The resulting ability to isolate ions from external perturbations is an ideal premise for precision spectroscopy, with no apparent fundamental limit for further reductions of the clock's uncertainties to the $10^{-19}$ range and beyond. Besides the application of improved frequency references, several fields of fundamental research can also benefit from these levels of precision, e.g.~the search for new physics beyond the standard model or possible violations of general relativity \cite{Ludlow2014,Huntemann2014a,Godun2014,Pruttivarasin2015,Dzuba2016,Delva2017}. Stable and reproducible optical clocks also pioneer interdisciplinary applications, such as chronometric leveling in geodesy~\cite{Bjerhammar1985, Vermeer1983, Mehlstaeubler2017}. Here, a frequency comparison of two remote clocks reveals their difference in a gravitational potential via the relativistic time dilation. Fractional frequency uncertainties of $10^{-18}$ and better will allow for the resolution of height differences below $\unit[1]{cm}$ in Earth's gravitational potential, enabling the use of optical clocks as highly sensitive quantum sensors. 

This work addresses a fundamental limitation in the current generation of optical ion clocks: the low signal-to-noise ratio of a single quantum absorber \cite{Itano1993}. With today's typical interrogation times $T_\mathrm{int}\approx\unit[100]{ms}$, the resulting statistical uncertainties are on the order of a few $10^{-15}/\sqrt{\tau}$, which means that averaging times $\tau$ of more than ten days are necessary to resolve the atomic transition frequency at the level of $10^{-18}$. State-of-the-art ion clock comparisons are limited by these long time scales. Simultaneous interrogation of $N$ ions \cite{Champenois2010,Herschbach2012,Arnold2015} already allows to resolve a specific frequency after $1/N$ of the averaging time needed with a single ion. With multi-ion spectroscopy, a new generation of ion clocks could be operated with a cascaded clock scheme, in which the local oscillator is successively stabilized, with increasing $T_\mathrm{int}$, to separate atomic ensembles \cite{Rosenband2013,Borregaard2013}. A further option is to reduce the measurement noise by employing non-classical collective states \cite{Leroux2010,Kessler2014,Lebedev2014}. So far, the major challenge hindering the implementation of a multi-ion clock was to extend the superb control over the trapping environment and ion motional dynamics to strongly coupled Coulomb crystals. These challenges are in part shared with efforts to advance quantum simulation with ions \cite{Schneider2012,Blatt2012,Hess2017} and quantum information processing \cite{Blatt2008} experiments in terms of system size and control.

In this work, we benchmark a precision ion trap platform for spectroscopy with chains of ions. We discuss and experimentally determine relevant sources of frequency uncertainty related to multi-ion operation in this geometry. Ion clocks have the advantage that all systematic frequency shifts and their contribution to the uncertainty budget can be determined via ``leverage'', in which either the shift itself is increased, or the underlying physical quantity is precisely measured via its effect on another observable \cite{Ludlow2014,Mehlstaeubler2017}. For the example of a clock based on multiple \ion{In} ions, sympathetically cooled with \ion{Yb}, we show that the overall uncertainty contribution related to multi-ion operation can be reduced to below $1\times 10^{-19}$. With our method, ion numbers around 100 are supported at this level of accuracy. Besides \ion{In} \cite{Becker2001,Ohtsubo2017}, other clock ion species that feature a transition with low sensitivity to electric field gradients, e.g.~\ion{Al} \cite{Beloy2017}, \ion{Yb} (${}^2$S${}_{1/2}\leftrightarrow {}^2$F${}_{7/2}$) \cite{Huntemann2016}, \ion{Lu} \cite{Arnold2015} and \ionm{229}{Th} \cite{Peik2003}, can also benefit from our approach.\\

This paper is organized as follows: Section \ref{experimental_setup} introduces the new ion trap platform and the experimental parameters for which systematic clock frequency shifts and their uncertainties are derived. In section \ref{E2shift_section}, we calculate the electric quadrupole shifts for ions within linear Coulomb crystals and derive the expected corresponding uncertainties. Time dilation and AC Stark shifts due to ion motion are discussed in section \ref{motional_shift_section}. Section \ref{further_shifts_section} treats further relevant shifts due to black-body radiation, background gas collisions and magnetic fields. We summarize our findings in section \ref{conclusion_section} by presenting a projected multi-ion related uncertainty budget.

\section{\label{experimental_setup}Experimental setup}
\begin{figure}
	\centerline{\includegraphics[width=.48\textwidth]{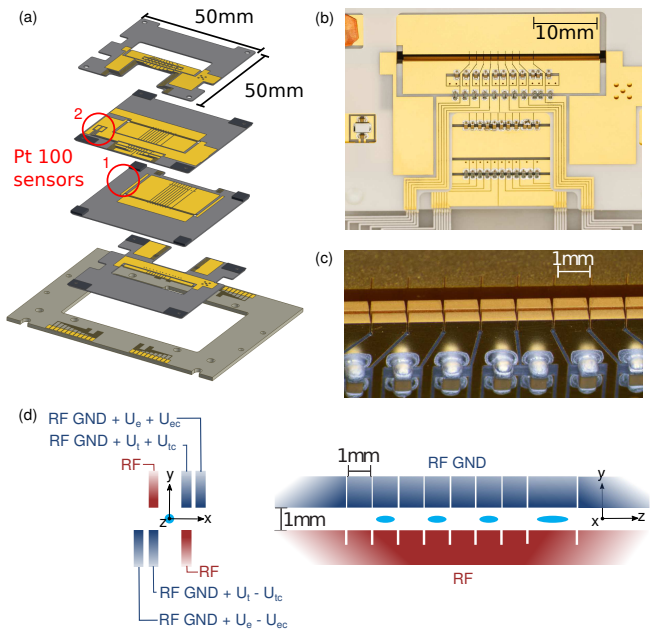}}
	\caption{\label{trapfigure}Scalable precision ion trap array. (a) Trap assembly from a stack of four wafers (taken from \cite{Keller2017}). (b) Photograph of the trap, showing the onboard low-pass filters and one thermistor. (c) Close-up on the trapping region. (d) Electrode geometry: the trap consists of one $\unit[2]{mm}$-long and seven $\unit[1]{mm}$-long segments. The rf electrodes are shown in red, and static voltages applied to the blue rf ground electrodes are used to confine the ions axially, compensate stray electric fields and control the radial mode frequency splitting and orientations.}
\end{figure}

The ion trap array presented in this work is formed by four wafers with gold electrodes, as shown in Fig.~\ref{trapfigure}a-c. It features an electrode design that was tested previously in a PCB prototype trap \cite{Pyka2014}. The $\unit[380]{\mu m}$ thick aluminum nitride wafers are laser cut and sputtered with $\unit[4]{\mu m}$ of gold, which is laser structured to form separate trapping segments. The assembly from monolithic wafers and the use of laser machining ensure scalability and symmetrically shaped electrodes with manufacturing tolerances below $\unit[10]{\mu m}$. To control the black-body radiation shift, temperature monitoring during clock operation is provided by two integrated Pt100 sensors on the innermost wafers, shown in Fig.~\ref{trapfigure}a,b.\\

RC filters with a cut-off frequency of $\unit[113]{Hz}$ are integrated on the trap wafers close to the DC control electrodes. The respective voltages are generated in digital-to-analog-converters (DACs), the outputs of which are scaled, combined, and low-pass filtered with a cut-off of $\unit[1]{kHz}$. Figure \ref{trapfigure}d summarizes all applied static voltages: $U_\mathrm{t,e}$ provide axial confinement and control over the orientation of the radial principal axes, whereas $U_\mathrm{tc,ec}$ are used to compensate stray electric fields. All of these voltages have RMS fluctuations below $\unit[100]{\mu V}$, with differential fluctuations of less than $\unit[30]{\mu V}$. Ions are loaded by photo-ionization from a thermal beam, which is collimated to a dedicated segment, avoiding contamination of the electrodes elsewhere.\\

\begin{figure}
	\centerline{\includegraphics[width=.48\textwidth]{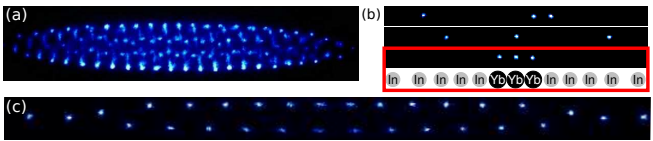}}
	\caption{\label{coulombcrystals} Coulomb Crystals. (a) Large Coulomb crystal (here: fluorescing \ion{Yb} ions), which can serve as a high-stability frequency reference. (b) Crystal configuration considered for spectroscopy with high accuracy (see Table \ref{uncertaintybudget}). The sketch corresponds to the lower of the three displayed crystal examples of linear \ion{In}/\ion{Yb} crystals. The axial extension is $\unit[219]{\mu m}$ ($\unit[61]{\mu m}$) for $\nu_\mathrm{ax}=\unit[30]{kHz}$ ($\nu_\mathrm{ax}=\unit[205]{kHz}$) (figure taken from \cite{Keller2017}). (c) Coulomb crystal with a symmetric topological defect (as used in \cite{Kiethe2017}).}
\end{figure}

The trap was developed to support simultaneous trapping of ion ensembles for precision spectroscopy \cite{Herschbach2012}. Large Coulomb crystals of hundreds of ions (Fig.~\ref{coulombcrystals}a) can serve as high-stability frequency references, while separate short ion chains of about 10 ions each can be used for high-accuracy clock interrogation (Fig.~\ref{coulombcrystals}b). The intrinsic symmetry of the trap has enabled the storage of symmetric Coulomb crystals with topological defects for the investigation of their dynamics and the observation of the phase transition between sticking and sliding regimes in atomic friction processes \cite{Kiethe2017} (Fig.~\ref{coulombcrystals}c).\\

The crystal configuration we consider in this work for clock operation consists of three \ion{Yb} cooling ions and ten \ion{In} clock ions and is shown in Fig.~\ref{coulombcrystals}b. Multiple of these ion chains can be simultaneously stored in neighboring trapping segments, forming an array of Coulomb crystals. Separation of the ion ensemble into such relatively short chains limits the complexity of the motional spectrum and allows internal state readout with single-ion resolution. We investigate two scenarios, of which configuration (\emph{A}) is optimized for the lowest systematic uncertainties, while configuration (\emph{B}) is simpler to implement experimentally. Both begin with sympathetic Doppler cooling on the ${}^2$S$_{1/2} \leftrightarrow {}^2$P${}_{1/2}$ transition in \ion{Yb} at $\unit[369.5]{nm}$ ($\Gamma=2\pi\times\unit[19.6]{MHz}$). In configuration (\emph{A}), a second cooling stage follows on the narrow ${}^1$S${}_0 \leftrightarrow {}^3$P${}_1$ intercombination line in \ionm{115}{In} at $\unit[230.6]{nm}$ ($\Gamma=2\pi\times\unit[360]{kHz}$). Configuration (\emph{B}) relies on clock interrogation at the \ion{Yb} Doppler temperature. Fluorescence from both species can be observed with an EMCCD camera and a photomultiplier tube (PMT). The respective confinement parameters are optimized for minimal overall systematic uncertainties, based on considerations detailed throughout the following sections. For sympathetic cooling on \ion{Yb}, a trap aspect ratio close to the phase transition from a linear chain to a two-dimensional crystal provides strong Coulomb coupling between the ions. The radial and axial trap frequencies of configuration (\emph{B}) are therefore $\nu_\mathrm{rad,In}=\unit[1.5]{MHz}$ and $\nu_\mathrm{ax,In}=\unit[205]{kHz}$, yielding a chain length of $l=\unit[61]{\mu m}$. Second-stage cooling with \ion{In} allows for lower temperatures and a weaker axial confinement. Configuration (\emph{A}) uses trap frequencies of $\nu_\mathrm{rad,In}=\unit[750]{kHz}$ and $\nu_\mathrm{ax,In}=\unit[30]{kHz}$, with a resulting chain length of $l=\unit[219]{\mu m}$ for the 13 ions. Clock spectroscopy is carried out on the  ${}^1$S${}_0 \leftrightarrow {}^3$P${}_0$ transition at $\unit[236.5]{nm}$ ($\Gamma=2\pi\times\unit[820]{mHz}$). In the following, we discuss systematic shifts of this transition during multi-ion operation.\\

\section{\label{E2shift_section}Electric quadrupole shift}
\begin{figure}
	\centerline{\includegraphics[width=.5\textwidth]{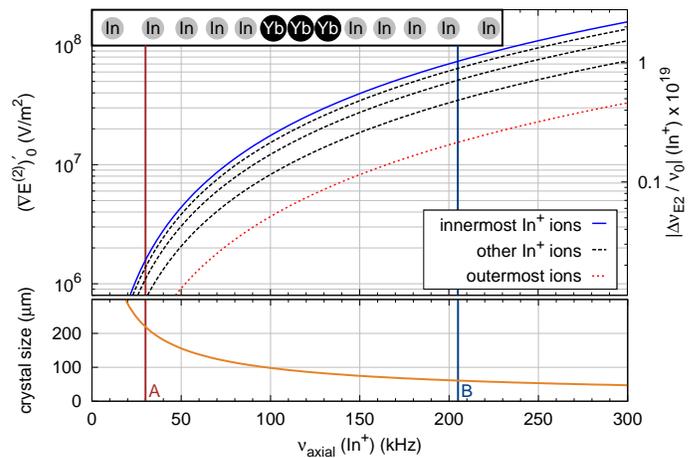}}
	\caption{\label{egradientfigure}Electric field gradients in a 13-ion chain as a function of the axial confinement. The key refers to the crystal configuration shown in the inset. The right hand scale shows the corresponding quadrupole shift for the transition from $^1\mathrm{S}_0\textnormal{,} m_F=\pm9/2$ to $^3\mathrm{P}_0\textnormal{,} m_{F^\prime}=\pm7/2$ in \ionm{115}{In}. The bottom graph shows the respective total chain length. Vertical bars indicate the parameters for an indium cooled (A) and sympathetically cooled (B) configuration, respectively.}
\end{figure}

The charges of neighboring ions within a Coulomb crystal can lead to considerable electric field gradients, which affect internal state energies by coupling to their electric quadrupole (E2) moments. The corresponding Hamiltonian has the form \cite{Itano2000}
\begin{equation}
\label{E2Hamiltonian}
H_{E2}=\nabla E^{(2)}\Theta^{(2)}\;\textnormal{,}
\end{equation}
where $\nabla E^{(2)}$ is the 2nd-rank field gradient tensor and $\Theta^{(2)}$ the electric quadrupole moment tensor of an electronic state. In the $^1\mathrm{S}_0\leftrightarrow{}^3\mathrm{P}_0$ transitions of two-electron systems, the rotational symmetry ($J=0$) of both clock states would ideally result in vanishing quadrupole moments. Hyperfine-mediated state mixing, however, leads to small finite E2 moments in the excited states \cite{Garstang1962}. Recent calculations of the resulting quadrupole moments for the respective states in \ion{In} yield a value of $\Theta({}^3\mathrm{P}_0,{}^{115}\mathrm{In}^+)=-1.6(3)\times10^{-5}ea_B^2$ \cite{Beloy2017}, where $a_B$ is the Bohr radius, $e$ is the elementary charge, and $\Theta$ is defined as the matrix element $\langle m_F=F,F\vert\Theta_0^{(2)}\vert m_F=F,F\rangle$ under the assumption that $H_{E2}$ can be treated as a perturbation with respect to the Zeeman Hamiltonian. Due to the small quadrupole moment, we neglect higher-order contributions from the rf electric field and only take the static part of the trap potential into account. Following the notation in \cite{Wuebbena2012,Beloy2017}, it can be approximated at the ion positions as
\begin{align}
\label{phi_dc}
\Phi_\mathrm{dc}=\frac{m\omega_\mathrm{ax}^2}{2e}\left(z^2-\alpha x^2-(1-\alpha)y^2\right)\\
\textnormal{with}\quad\omega_\mathrm{ax}=\sqrt{\frac{2e\kappa_\mathrm{ax}U_\mathrm{ax}}{mz_0^2}}\nonumber\\
\textnormal{and}\quad\alpha=\frac{1}{2}\left(1-\frac{\kappa_tU_t}{d_0^2}\frac{2e}{m\omega_\mathrm{ax}^2}\right)\;\textnormal{,}\nonumber
\end{align}
where $m$ denotes the ion mass, $\omega_\mathrm{ax}$ the axial secular frequency (note that $m\omega_\mathrm{ax}^2$ is mass independent), $d_0$ and $z_0$ the distance from the segment center to the quadrupole electrodes and the neighboring segments, respectively, $U_\mathrm{ax}$ the axial trapping voltage, and the $\kappa$ factors are geometrical corrections of order unity. For clarity, we have omitted the contribution of $U_\mathrm{e}$, which consists of a radially rotated version of the potential produced by $U_\mathrm{t}$ and can be accounted for by redefining the $x$ and $y$ directions and adjusting $\alpha$. The corresponding static electric field gradients are
\begin{align}
\frac{\partial \vec{E}_{\mathrm{trap}}}{\partial x,y}&=\frac{m\omega_\mathrm{ax}^2}{2e}\left(1\pm(2\alpha-1)\right)\vec{e}_{x,y}\;\textnormal{,}\\
\frac{\partial \vec{E}_{\mathrm{trap}}}{\partial z}&=-\frac{m\omega_\mathrm{ax}^2}{e}\vec{e}_z\;\textnormal{.}\nonumber
\end{align}
The axial ($z$) component of the contribution from neighboring ions at the position of ion $i$ takes the form
\begin{equation}
\frac{\partial \vec{E}_{i,\mathrm{ions}}}{\partial z}=-\frac{2m\omega_\mathrm{ax}^2}{e}\sum_{j\neq i}\frac{1}{\vert u_i-u_j\vert^3}\vec{e}_z\;\textnormal{,}
\end{equation}
with the scaled equilibrium positions \cite{James1998} $u_i=z_i/l$, 
\begin{equation}
l^3=\frac{e^2}{4\pi\varepsilon_0m\omega_\mathrm{ax}^2}\;\textnormal{.}
\end{equation}
Due to the Laplace condition and rotational symmetry,
\begin{equation}
\frac{\partial E_{i,\mathrm{ions},x}}{\partial x}=\frac{\partial E_{i,\mathrm{ions},y}}{\partial y}=-\frac{1}{2}\frac{\partial E_{i,\mathrm{ions},z}}{\partial z}\;\textnormal{.}
\end{equation}
In total, the components of $(\nabla E_i^{(2)})$ in the spherical basis oriented with respect to the trap axis are \cite{Itano2000}
\begin{align}
(\nabla E_i^{(2)})_0&=-\frac{1}{2}\frac{\partial E_{z}}{\partial z}=\frac{m\omega^2_\mathrm{ax}}{e}\left(\frac{1}{2}+\sum_{j\neq i}\frac{1}{\vert u_i-u_j\vert^3}\right)\nonumber\\
(\nabla E_i^{(2)})_{\pm1}&=\pm\frac{1}{\sqrt{6}}\left(\frac{\partial}{\partial x}\pm i\frac{\partial}{\partial y}\right)E_{z}=0\\
(\nabla E_i^{(2)})_{\pm2}&=-\frac{1}{2\sqrt{6}}\left(\frac{\partial}{\partial x}\pm i\frac{\partial}{\partial y}\right)(E_{x}\pm iE_{y})\nonumber\\
&=\frac{1}{\sqrt{6}}\frac{m\omega^2_\mathrm{ax}}{e}\left(\frac{1}{2}-\alpha\right)\;\textnormal{.}\nonumber
\end{align}
In order to calculate the scalar product of (\ref{E2Hamiltonian}), the final step is to find $(\nabla E_i^{(2)})^\prime_0$ in the coordinate system of the Zeeman eigenstates, in which the $z^\prime$ axis is parallel to $\vec{B}$. This is achieved with a coordinate transformation that consists of a rotation by $\phi$ around $z$, followed by a $\theta$ rotation around the new $y^{\prime\prime}$ axis \cite{Devanathan2002}:
\begin{align}
\label{gradE_tensor}
(\nabla E_i^{(2)})^\prime_0&=\frac{m\omega_\mathrm{ax}^2}{e}\left[\frac{\sin^2\theta\cos2\phi}{2}\left(\frac{1}{2}-\alpha\right)\phantom{\sum_{j\neq i}}\right.\\\nonumber
&\left.+\frac{3\cos^2\theta-1}{2}\left(\frac{1}{2}+\sum_{j\neq i}\frac{1}{\vert u_i-u_j\vert^3}\right)\right]\;\textnormal{.}
\end{align}
Since $\Theta^{(2)}$ is defined for the stretched states $\vert m_F\vert=F$, its value for other $m_F$ states is scaled by a ratio of Clebsch-Gordan coefficients, which can be expressed as \cite{Beloy2017}
\begin{equation}
\label{e2momentmdependence}
\Theta(F,m_F)=\frac{3m_F^2-F(F+1)}{F(2F-1)}\Theta(F,m_F=F)\;\textnormal{.}
\end{equation}
For the $m_{F^\prime}=7/2$ substate, chosen to minimize the linear Zeeman shift (cf.~section \ref{magnetic_field}), the prefactor takes the value $1/3$. The experimental challenges for reducing the quadrupole shift uncertainty depend on the orientation of $\vec{B}$: for the choice of $\theta=0$, i.e.~$\vec{B}$ parallel to the trap axis, the first term in (\ref{gradE_tensor}) vanishes, while the second term is first-order insensitive to fluctuations in $\theta$. This only leaves a first-order dependence on $\omega_\mathrm{ax}$, the experimental stabilization of which is the least challenging among these parameters. Alternatively, (\ref{gradE_tensor}) could be made to vanish entirely, e.g.~by setting $\theta=\cos^{-1}(1/\sqrt{3})\approx\unit[55]{{}^\circ}$ and $\phi=\pi/4$, at the expense of a first-order sensitivity to these angles. Figure \ref{egradientfigure} shows $(\nabla E_i^{(2)})^\prime_0$ for $\theta=0$ as a function of the single \ion{In} ion axial trap frequency $\nu_\mathrm{ax}=\omega_\mathrm{ax}/(2\pi)$ at the clock ion positions in the example crystal configuration. For the axial confinements of $\unit[30]{kHz}$ (configuration \emph{A}) and $\unit[205]{kHz}$ (configuration \emph{B}), the highest fractional shifts are $-0.02\times10^{-19}$ and $-1.1\times10^{-19}$, respectively, such that a relative axial frequency uncertainty of $10^{-2}$ is sufficient to yield uncertainties close to $1\times10^{-21}$. The \ion{In} cooled configuration (\emph{A}) benefits from its weaker confinement and the resulting increase in ion distances.
Finally, we note that several cancellation schemes exist for the electric quadrupole shift, which can be used for species with higher intrinsic sensitivity, such as \ion{Lu} or \ion{Yb}. They include averaging over the full Zeeman structure \cite{Dube2005}, averaging over the hyperfine structure \cite{Barrett2015}, or averaging over three mutually orthogonal directions of $\vec{B}$ \cite{Itano2000}.

\section{\label{motional_shift_section}Motional effects}
\subsection{Excess micromotion}
\begin{figure}
	\centerline{\includegraphics[width=0.45\textwidth]{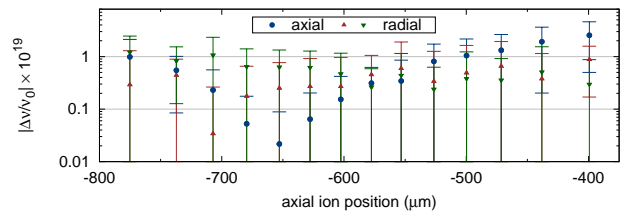}}
	\caption{\label{IIEMM} Time dilation shift due to excess micromotion (from \cite{Keller2017}). A 14-ion \ionm{172}{Yb} crystal was used to probe rf electric amplitudes along the line of minimal potential in the 2-mm segment. The graph shows the corresponding time dilation shifts for \ionm{115}{In} ions.}
\end{figure}

Excess micromotion (EMM) is the driven motion of an ion at the frequency $\Omega_\mathrm{rf}$ when the confining field has a nonzero amplitude at the equilibrium position. Its adverse effects include time dilation and AC Stark shifts \cite{Berkeland1998} (which can be made to compensate each other in transitions with negative differential polarizabilities \cite{Madej2012}), as well as increased heating \cite{Blakestad2009}. Our scheme avoids the latter, as well as the restriction to species with negative differential polarizabilities, by storing all ions at positions of vanishing rf electric fields \cite{Herschbach2012}. Using in-situ EMM amplitude measurements in \ionm{172}{Yb} crystals \cite{Keller2017}, we have demonstrated the ability to store extended ion chains of up to $\unit[400]{\mu m}$ ($\unit[2]{mm}$) length with time dilation shifts close to $1\times10^{-19}$ and below $10^{-18}$, respectively. Fig.~\ref{IIEMM} shows the result of such a measurement for the 2mm-long trapping segment. The axial component of these measurements was stable over the course of more than 6 months, such that an initial characterization is sufficient. The radial contributions were homogeneous over the whole chain and can therefore be compensated during clock operation by monitoring the fluorescence of a single ion with a dedicated PMT during the Doppler cooling stage.

\subsection{Thermal motion}
We determine the motional mode eigenvectors from a 2nd-order expansion of the potential energy with respect to mass-weighted ion displacements from their equilibrium positions as described, e.g.~in \cite{Morigi2001,Home2013}. The eigenmode frequencies are expressed as $\omega_\alpha$, and the eigenvector component of ion $i$ for mode $\alpha$ in mass-weighted space is denoted by $\beta_{\alpha,i}^\prime$. These vectors are normalized such that $\sum_i\beta_{\alpha,i}^\prime=1$. The displacement of ion $i$ can then be expressed in terms of the motional mode excitations as
\begin{equation}
r_i=\frac{1}{{\sqrt{m_i}}}\sum_\alpha\beta_{\alpha,i}^\prime\pi_\alpha\;\textnormal{,}
\end{equation}
where $\pi_\alpha$ is the effective position of the mode $\alpha$. For a classical oscillator with thermal excitation $k_BT_\alpha$, it takes the form
\begin{equation}
\pi_\alpha=\sqrt{\frac{2k_BT_\alpha}{\omega_\alpha^2}}\cos(\omega_\alpha t+\varphi_\alpha)\;\textnormal{.}
\end{equation}
Alternatively, it can be expressed as a position operator using the mode annihilation operator $\hat{a}_\alpha$:
\begin{equation}
\hat{\pi}_\alpha=\sqrt{\frac{\hbar}{2\omega_\alpha}}\left(\hat{a}_\alpha+\hat{a}^\dagger_\alpha\right)\;\textnormal{.}
\end{equation}

The equilibrium temperature for each eigenmode is determined by equating the respective heating and cooling rates,
\begin{equation}
\dot{E}_\mathrm{cool,laser,\alpha}+\dot{E}_\mathrm{heat,laser,\alpha}+\dot{E}_\mathrm{heat,ext,\alpha}=0\;\textnormal{.}
\end{equation}
An absorption event by ion $i$ changes the total kinetic energy by
\begin{align}
\Delta E_\mathrm{abs}&=\frac{1}{2m_i}\left(\left(p_i+\hbar k\right)^2-p_i^2\right)\\
&=\frac{(\hbar k)^2}{2m_i}+\hbar k\sum_\alpha\frac{\beta_{\alpha,i}^\prime}{\sqrt{m_i}}\dot{\pi}_\alpha\;\textnormal{,}\nonumber
\end{align}
where $p_i=m_i\dot{r}_i$ and $k$ is the projection of $\vec{k}$ onto the ion momentum. We linearize the scattering rate with respect to the ion velocity $\Gamma_\mathrm{sc}\approx\Gamma_{\mathrm{sc},0}(1+\vec{\rho} \dot{\vec{r}}_i)$ \cite{Leibfried2003} ($\vec{\rho}\,\Vert\,\vec{k}$). The overall rate of change for the energy can be obtained by averaging $\Gamma_\mathrm{sc}(\Delta E_\mathrm{abs}+\Delta E_\mathrm{em})$ over the ion trajectory, with $\Delta E_\mathrm{em}$ accounting for the spontaneous re-emission of photons, which is isotropic after summing over all polarizations. Since the motion of the individual modes is uncorrelated, they can be treated independently when averaging over typical Doppler cooling times of several $\unit{ms}$, as they are much longer than the respective oscillation periods. 
In total, we obtain for mode $\alpha$
\begin{align}
\label{laser_energy_rates}
\left\langle\dot{E}_{\alpha}\right\rangle=\Gamma_{\mathrm{sc},0}\sum_i&\left(\frac{\left(\beta_{\alpha,i}^\prime\hbar k_\alpha\right)^2}{2m_i}+\frac{\left(\beta_{\alpha,i}^\prime\hbar\frac{k}{\sqrt{3}}\right)^2}{2m_i}\right.\nonumber\\
+&\left.\vphantom{\frac{\left(\hbar (k_\alpha+\frac{k}{\sqrt{3}})\beta_{\alpha,i}^\prime\right)^2}{2m_i}}\frac{1}{2}\hbar k_\alpha\frac{\beta_{\alpha,i}^{\prime2}}{m_i}\rho_\alpha \dot{\pi}_{0,\alpha}^2\right)\;\textnormal{,}
\end{align}
with the thermal oscillator peak velocity $\dot{\pi}_{0,\alpha}=\sqrt{2k_BT_\alpha}$ and the projection $k_\alpha$ of $\vec{k}$ onto the mode principal axis. The sum in Eq.~\ref{laser_energy_rates} includes all cooling ions, and the first two terms constitute $\dot{E}_{\mathrm{heat,laser},\alpha}$, which accounts for heating due to velocity-independent scattering events and $\Delta E_\mathrm{em}$. Setting $\langle\dot{E}_\alpha\rangle+\dot{E}_\mathrm{heat,ext,\alpha}=0$ and solving for $T_\alpha$, we obtain the equilibrium temperature
\begin{equation}
T_\alpha=-\frac{\dot{E}_{\mathrm{heat,ext},\alpha}+\Gamma_{\mathrm{sc},0}\frac{1}{2}
\left((\hbar k_\alpha)^2+\frac{(\hbar k)^2}{3}\right)\sum_i\frac{\beta_{\alpha,i}^{\prime 2}}{m_i}}{k_B\Gamma_{\mathrm{sc},0}\rho_\alpha\hbar k_\alpha\sum_i\frac{\beta_{\alpha,i}^{\prime 2}}{m_i}}\;\textnormal{.}
\end{equation}
In all calculations below, we assume a cooling laser orientation $\vec{k}$ with equal projections onto the three principal axes.

\begin{figure}
  \centerline{\includegraphics[width=.45\textwidth]{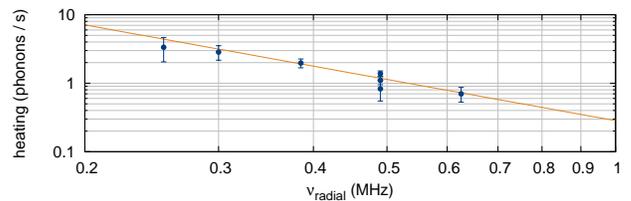}}
  \caption{\label{heatingratefigure}Frequency dependence of the radial heating rates, determined using a single \ionm{172}{Yb} ion cooled to the ground state. The fitted line corresponds to $\dot{\bar{n}}=\unit[2.8(2)\times10^{11}]{s^{-1}Hz^2}/\nu_\mathrm{radial}^2$.}
\end{figure}
In order to determine the expected external heating rates $\dot{E}_{\mathrm{heat,ext},\alpha}$, we measure the electric field power spectral density (PSD) $S_E$ in the $\unit[2]{mm}$-long segment by using a single \ion{Yb} ion as a field probe. Figure \ref{heatingratefigure} shows the measured heating rates $\dot{\bar{n}}=\dot{E}/(\hbar\omega)$ as a function of frequency, from which the PSD can be obtained via \cite{Turchette2000}
\begin{equation}
S_E(f=\nu)=\frac{4mhf\dot{\bar{n}}(\nu)}{e^2}\;\textnormal{.}
\end{equation}
We observe a frequency scaling of $\dot{\bar{n}}(\nu)\propto\nu^{-1.8(3)}$, similar to observations in other ion traps (e.g.~\cite{Bruzewicz2015,Boldin2017}). A fit with the exponent fixed at $-2$ results in $\dot{\bar{n}}=\unit[2.8(2)\times10^{11}]{s^{-1}Hz^2}/\nu^2$, corresponding to $S_E=\unit[8.4(5)\times10^{-9}]{(V/m)^2/f}$. We attribute this primarily to anomalous heating \cite{Brownnutt2014}, as estimates of other electric field noise sources are orders of magnitude lower: all DC voltages are low-pass filtered with a cutoff frequency of $\unit[113]{Hz}$ directly on the trap wafers. From voltage noise measurements of the DC trap voltage supplies, we expect a contribution of $\unit[2.2\times10^{-18}]{(V/m)^2/Hz}$ (ca.~$2\times10^{-4}$ phonons per second for \ionm{115}{In}) at $\unit[500]{kHz}$, whereas Johnson-Nyquist noise is expected to contribute $\unit[3\times10^{-17}]{(V/m)^2/Hz}$ ($3\times10^{-3}$ phonons per second). These values also show that the on-board filter cutoff frequency could be increased by two orders of magnitude without an appreciable effect on ion heating, which would allow faster changes to the confinement. As the distances to the electrodes are large compared to the crystal extension, we approximate the fields responsible for heating as uniform across the chain. The heating rates of higher-order modes can then be calculated as \cite{Kielpinski2000}
\begin{equation}
\dot{\bar{n}}_\alpha=\left(\sum_i\frac{\beta_{\alpha,i}^\prime}{\sqrt{m_i}}\right)^2\frac{e^2}{4\hbar\omega_\alpha}S_E(\omega_\alpha)\;\textnormal{.}
\end{equation}
With these calculations, equilibrium temperatures below $\unit[0.7]{mK}$ are obtained for sympathetic cooling of all modes in configuration (\emph{B}) as introduced in section \ref{experimental_setup}. Of these temperatures, only those of the in-phase modes (radial and axial) and the next higher-order even symmetry radial modes are determined by the external heating rates, while all other modes are calculated to be essentially at the ideal Doppler temperature. We note that higher-order spatial variations of the electric fields responsible for heating could adversely affect these results, in particular for modes with odd symmetry, and need further experimental investigation.

\subsubsection{Thermal time dilation shift}
\begin{figure}
  \centerline{\includegraphics[width=.45\textwidth]{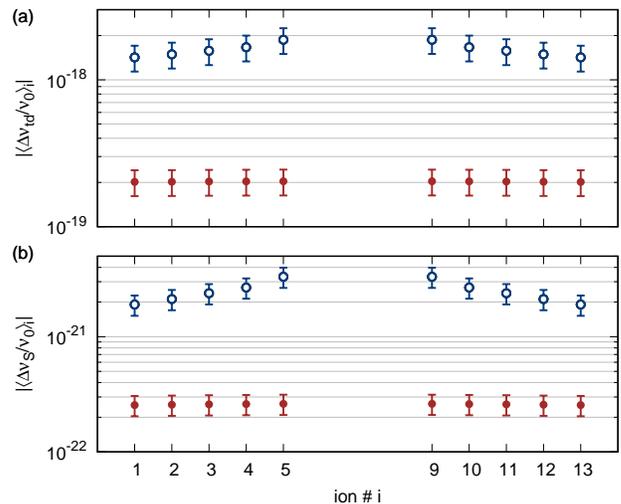}}
  \caption{\label{IMM_figure}Fractional frequency shifts of each clock ion (cf.~Fig.~\ref{coulombcrystals}b) due to thermal motion. Closed symbols depict the results for configuration (\emph{A}) (cooling on the narrow intercombination line in \ion{In}); open symbols represent configuration (\emph{B}) (sympathetic cooling on \ion{Yb} only). (a) Time dilation shift due to secular motion and intrinsic micromotion for all $3N$ modes. (b) AC Stark shift due to intrinsic micromotion.}
\end{figure}

The fractional frequency shift induced by time dilation in the moving ion frame of reference can be expressed as the ratio between kinetic energy and rest energy:
\begin{equation}
\left\langle\frac{\Delta\nu_\mathrm{td}}{\nu_0}\right\rangle=-\frac{\langle v^2\rangle}{2c^2}\;\textnormal{.}
\end{equation}
Besides the secular motion itself, $v$ contains an intrinsic micromotion contribution for radial modes. For a single ion in a purely ponderomotive potential, both contributions are approximately equal (remarkably, this addition is also included in a derivation via the mass defect due to excitation of the clock transition and its influence on the ponderomotive confinement, see \cite{Yudin2017} and Appendix \ref{massdefect}). Static potential terms, however, can allow for excursions into regions of higher rf field for a given energy \cite{Wuebbena2012}. These contributions arise from the radial repulsion caused by the axial confinement, the induced radial anisotropy, and the repulsion by neighboring ions. The average velocities for thermal mode excitations can be derived from the classical ion trajectory \cite{Major2005},
\begin{align}
\label{classicalionmotion}
x_i(t)=\sum_\alpha x_{0,i,\alpha}\cos(\omega_\alpha t)\left(1+\frac{q_i}{2}\cos(\Omega_\mathrm{rf}t)\right)\\
\textnormal{with}\quad x_{0,i,\alpha}=\beta_{\alpha,i}^\prime\sqrt{\frac{2k_BT_\alpha}{m_i\omega_\alpha^2}}\;
\end{align}
and the Mathieu $q$-parameter \cite{Paul1990} in the respective radial direction,
\begin{equation}
q_{i,x/y}=\pm\frac{2e\kappa_\mathrm{rf}U_\mathrm{rf}}{m_id_0^2\Omega_\mathrm{rf}^2}\;\textnormal{.}\label{mathieuq}
\end{equation}
Differentiating and averaging with respect to time, we obtain
\begin{equation}
\langle v^2_i\rangle =\sum_\alpha\langle v_{i,\alpha}^2\rangle=\sum_\alpha x_{0,i,\alpha}^2\left(\frac{\omega_\alpha^2}{2}+\frac{q_i^2}{16}\left(\Omega_\mathrm{rf}^2+\omega_\alpha^2\right)\right)\;\textnormal{,}
\end{equation}
where the first step assumes sufficiently long averaging times to treat the motion of different modes as uncorrelated. The full expression for the thermal time dilation shift is then
\begin{equation}
\label{immD2shift}
 \left\langle\frac{\Delta\nu_\mathrm{td}}{\nu_0}\right\rangle_i=-\sum_\alpha\frac{k_BT_\alpha}{2c^2}\frac{\beta_{\alpha,i}^{\prime2}}{m_i}\left(1+\frac{\omega_{\mathrm{pm},i}^2}{\omega_\alpha^2}+\frac{\omega_{\mathrm{pm},i}^2}{\Omega_\mathrm{rf}^2}\right)\;\textnormal{,}
\end{equation}
with the purely ponderomotive trap frequency $\omega_\mathrm{pm,i}=\vert q_i\vert\Omega_\mathrm{rf}/\sqrt{8}$.
While the third term in Eq.~(\ref{immD2shift}) is typically negligible, the second term increases with decreasing radial mode frequencies, e.g.~when the Coulomb repulsion contributes more strongly to the overall potential. Weak axial confinement is therefore favorable for reducing this shift. Figure \ref{IMM_figure}a shows the calculated shift for each clock ion in cooling configurations (\emph{A}) and (\emph{B}).

\subsubsection{Thermal AC Stark shift}
In addition to micromotion, the rf electric field also induces an AC Stark shift,
\begin{equation}
\left\langle\frac{\Delta\nu_S}{\nu_0}\right\rangle=-\frac{\Delta\alpha_\mathrm{stat}}{h\nu_0}\frac{\langle E_\mathrm{rf}^2\rangle}{2}\;\textnormal{,}
\end{equation}
which depends on the static differential polarizability of the clock states $\Delta\alpha_\mathrm{stat}$, since $\Omega_\mathrm{rf}$ is far-detuned from any strongly allowed electronic transitions. The magnitude of $\langle E_\mathrm{rf}^2\rangle$ can be derived by dropping the secular motion terms from the expression (\ref{classicalionmotion}) and using the fact that $eE_\mathrm{rf}$ is the force responsible for micromotion, $E_\mathrm{rf}=m_i\ddot{x}_{i,\mathrm{mm}}/e$:
\begin{align}
&\left\langle\frac{\Delta\nu_S}{\nu_0}\right\rangle_i=-\frac{\Delta\alpha_\mathrm{stat}}{h\nu_0}\frac{m_i^2}{2e^2}\langle \ddot{x}_{i,\mathrm{mm}}^2\rangle\\
&=-\frac{\Delta\alpha_\mathrm{stat}}{h\nu_0}\frac{m_i}{2e^2}\sum_\alpha\beta_{\alpha,i}^{\prime 2}k_BT_\alpha\omega_{\mathrm{pm},i}^2\left(\frac{\Omega_\mathrm{rf}^2}{\omega_\alpha^2}+\frac{\omega_\alpha^2}{\Omega_\mathrm{rf}^2}+6\right)\;\textnormal{,}\nonumber
\end{align}
where, as before, we assumed sufficiently long averaging times to neglect correlations in the contributions from separate modes. As for the time dilation shift, a decrease in Coulomb coupling, i.e.~weak axial confinement, helps to reduce this contribution. The expected shifts for the discussed crystal configurations are shown in Fig.~\ref{IMM_figure}b.\\

When relying on sympathetic cooling, both of these thermal shifts lead to conflicting requirements for the axial confinement: high trap frequencies lead to strong Coulomb coupling and thus higher cooling rates, while at the same time increasing the motional shifts at a given temperature. The parameters of configuration (\emph{B}) have been chosen as a trade-off between these conditions. This conflict could be circumvented by adiabatically switching between separate cooling and interrogation configurations.

\subsection{Debye-Waller effect}
\begin{figure}
  \centerline{\includegraphics[width=.45\textwidth]{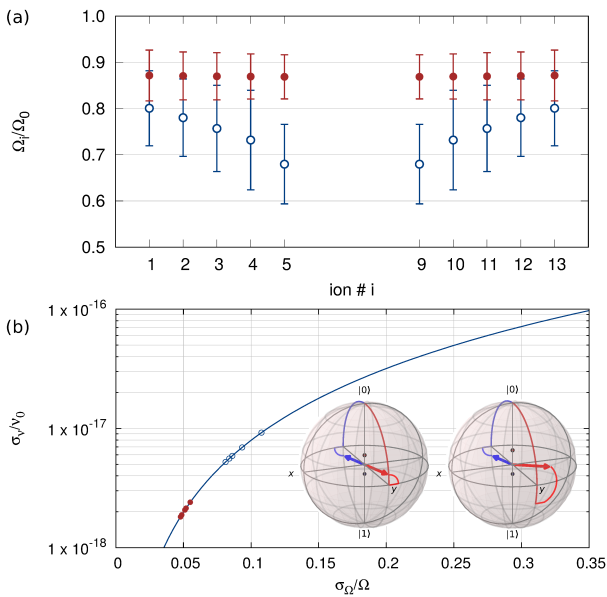}}\caption{\label{debyewallerramseyfigure}Debye-Waller effect in clock interrogation. (a) Mean Rabi frequencies and RMS fluctuations, normalized to the Rabi frequency of a free atom, for configuration (\emph{A}) (closed circles) and (\emph{B}) (open circles). (b) Effective broadening of the interrogation laser due the Rabi frequency fluctuations between interrogations in Ramsey spectroscopy with $T_\mathrm{int}=\unit[195]{ms}$, as predicted by Eq.~\ref{dwefflaserbroadening}. Closed and open symbols identify the respective impact of the RMS spreads shown in (a). The Bloch spheres in the inset illustrate the effect of a Rabi frequency mismatch on the error signal: on the left, the same Rabi frequency is assumed for both interrogations, while on the right, the Rabi frequency of the red (front) trace is increased by $\unit[20]{\%}$, leading to a higher difference in observed populations.}
\end{figure}

Besides its effect on accuracy, ion motion also influences the clock instability via the Debye-Waller effect \cite{Wineland1998}, which leads to a reduction and fluctuation of the clock laser Rabi frequency $\Omega_i$ for ion $i$. The fractional RMS fluctuations between experiments are \cite{Wineland1998}
\begin{equation}
\frac{\sigma_{\Omega,i}}{\Omega_{i}}=\sqrt{\left[\prod_\alpha I_0\left(2\eta_{\alpha,i}^2\sqrt{\bar{n}_\alpha(\bar{n}_\alpha+1)}\right)\right]-1}\,\textnormal{,}
\label{dwrms}
\end{equation}
where $I_0$ denotes the zeroth modified Bessel function, and the Lamb-Dicke parameter $\eta_{\alpha,i}$ of ion $i$ and mode $\alpha$ is given by
\begin{equation}
\eta_{\alpha,i}=k_\alpha\beta_{\alpha,i}^\prime\sqrt{\frac{\hbar}{2m_i\omega_\alpha}}\;\textnormal{,}
\label{lambdicke}
\end{equation}
for which again $k_\alpha$ is the projection of $\vec{k}$ onto the mode principal axis. The contribution of each mode increases with the spatial extent of the corresponding wavefunction, i.e.~with higher thermal excitation and with weaker confinement. Low-frequency modes, e.g.~in the complex spectrum of a large crystal, are therefore particularly critical. Since linear crystals require a weak axial confinement, clock spectroscopy is implemented along one of the radial principal axes. Figure \ref{debyewallerramseyfigure}a shows the calculated Rabi frequencies $\Omega_i$ of all clock ions, normalized to the Rabi frequency $\Omega_0$ of a free atom, as well as the RMS fluctuations $\sigma_{\Omega,i}$ between experiments for configurations (\emph{A}) and (\emph{B}). For sufficiently stable temperatures, the differences in the mean Rabi frequencies behave like spatial variations in the clock laser intensity. The resulting pulse area error can be corrected for in the calculation of frequency corrections by the clock servo as long as $\sigma_\Omega\ll\Omega$. The fluctuations between experiments, however, reduce the amount of obtainable information and therefore increase the clock instability. The Rabi interrogation scheme, in which the clock laser is applied during the full interrogation time, is first-order sensitive to these fluctuations and thus not feasible for our parameters. We therefore assume Ramsey interrogation. The error signal $S$ is typically derived by subtraction of the populations after two interrogation cycles with phase shifts of $\pm\pi/2$ between the laser pulses, and has the following form \cite{Ramsey1951}:
\begin{align}
\label{ramseysignal}
S=&\sin^2(\Omega_1 t)\cos^2\left(\frac{\Delta T_\mathrm{int}}{2}-\frac{\pi}{4}\right)\\\nonumber
-&\sin^2(\Omega_2 t)\cos^2\left(\frac{\Delta T_\mathrm{int}}{2}+\frac{\pi}{4}\right)\;\textnormal{,}
\end{align}
where $\Delta$ denotes the laser detuning from resonance. Assuming that the Rabi frequency of the second interrogation $\Omega_2$ differs from $\Omega_1$ by a fraction $e_\Omega=(\Omega_2-\Omega_1)/\Omega_1$, we obtain a frequency error of
\begin{align}
\varepsilon_\omega=&(S_{\Omega_2\neq\Omega_1}-S_{\Omega_2=\Omega_1})/(\partial S/\partial\Delta)\label{dwefflaserbroadening}\\\nonumber
=&\left[\left(\frac{\pi^2}{4}\cos^2\left(\frac{\Delta T_\mathrm{int}}{2}+\frac{\pi}{4}\right)\right)e_\Omega^2+\mathcal{O}(e_\Omega^3)\right]/T_\mathrm{int}\\\nonumber
\approx&\frac{\pi^2}{8T_\mathrm{int}}e_\Omega^2\;\textnormal{,}
\end{align}
where the last step assumed $\Delta\approx0$, as is typical for clock operation. The effect is similar to an increase in clock laser frequency instability, as Fig.~\ref{debyewallerramseyfigure}b shows for the ideal interrogation time of $T_\mathrm{int}=1/\Gamma=\unit[195]{ms}$ \cite{Peik2006} for \ion{In} (assuming $e_\Omega=\sigma_\Omega/\Omega$). The Bloch sphere insets visualize the effect: the two trajectories correspond to two measurements with opposite phase shifts of $\pm\pi/2$ used to obtain Eq.~\ref{ramseysignal}, assuming a small detuning $\Delta$ from resonance for clarity. In the left example, equal and optimal Rabi frequencies $\Omega_1=\Omega_2$ are assumed for both sequences, resulting in a symmetric deviation of the measured populations from $0.5$. The right example assumes $\Omega_2=1.2\Omega_1$, leading to an increased difference in populations, which would cause an overcorrection by the clock servo. In our case, even for the highest expected fluctuations of $\sigma_\Omega<0.11\Omega_0$ in configuration (\emph{B}), the effect is more than an order of magnitude below the quantum projection noise of $100$ \ion{In} ions. It can, however, quickly become relevant for stronger axial confinement or higher ion numbers, both of which further reduce the frequencies of the higher-order radial modes, or in three-dimensional crystals with low-frequency motional modes. More advanced pulse sequences, such as broadband composite pulses \cite{Wimperis1994} could be used to mitigate this issue at the expense of a lower interrogation duty cycle.

\section{\label{further_shifts_section}Further shifts}
\subsection{\label{BBR_section}Trap heating and BBR shift}
\begin{figure}
	\centerline{\includegraphics[width=.45\textwidth]{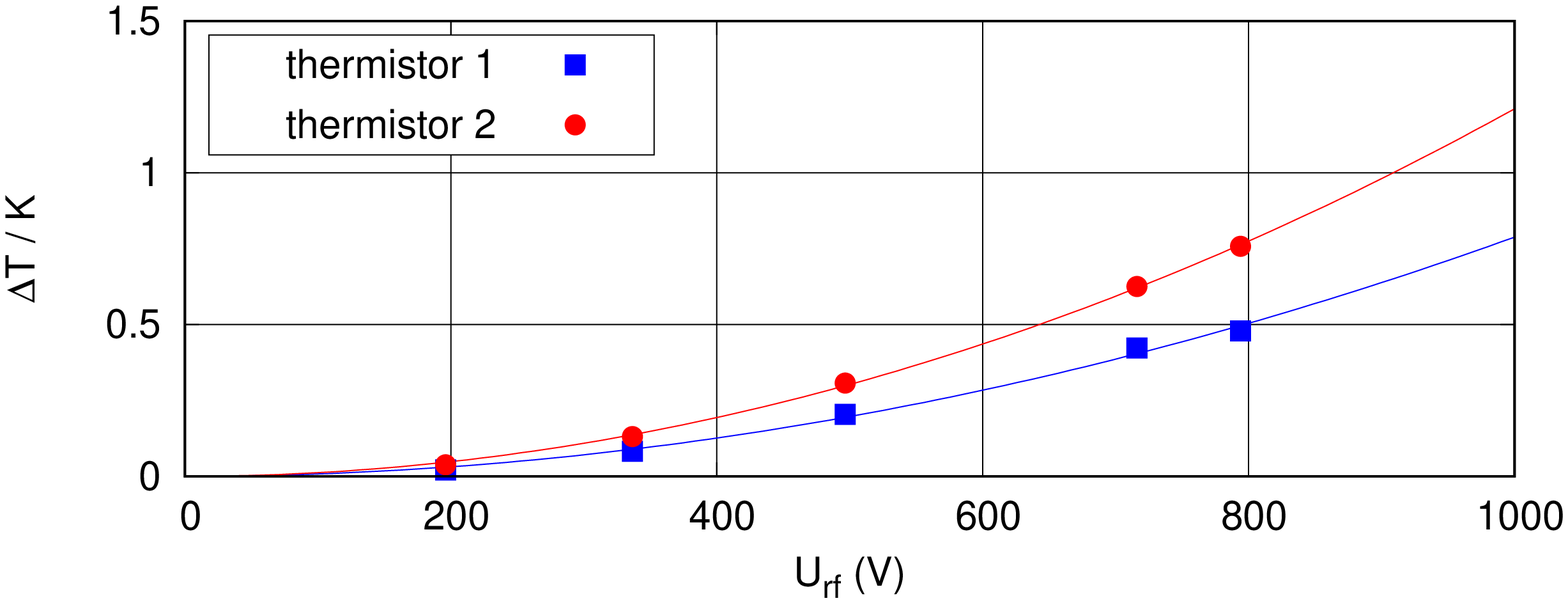}}
	\caption{\label{trapheatingfigure} Trap temperature increase at the two integrated Pt100 sensors as a function of the rf voltage amplitude. "Thermistor~1" refers to the sensor on the rf wafer closest to the carrier board. "Thermistor~2" is on the other rf wafer. The mean temperature increase of the trap for an rf voltage amplitude of $\unit[750]{V}$ ($\unit[1.5]{kV}$) at $\Omega_\mathrm{rf}=2\pi\times\unit[24.4]{MHz}$ is $\unit[0.6]{K}$ ($\unit[2.2]{K}$). The voltage uncertainty of each data point is less than $\unit[10]{V}$ and temperature rise uncertainties are around $\unit[15]{mK}$ and originate from fit errors of the experimental data. The lines are quadratic fits to the data points.}
\end{figure}

The AC Stark shift due to black-body radiation (BBR) can be expressed as \cite{Porsev2006}
\begin{equation}
h\Delta\nu_\mathrm{BBR}=-\frac{1}{2}\Delta\alpha_\mathrm{stat}(1+\eta)(\unitfrac[831.9]{V}{m})^2\left(\frac{T}{\unit[300]{K}}\right)^4\;\textnormal{,}
\end{equation}
where the dynamical correction $\eta$ can be neglected in the case of \ion{In}, for which all electric dipole allowed transitions coupling to the clock levels are in the UV, and thus far detuned from the room-temperature BBR spectrum. 
The two integrated Pt100 thermistors, shown in Fig.~\ref{trapfigure}, have been calibrated to within $\unit[70]{mK}$ \cite{Didier2018}. Figure \ref{trapheatingfigure} shows the observed temperature rise as a function of the applied rf voltage. The temperature measured with thermistor~2 is slightly higher due to its increased distance from to the carrier board, through which the trap is thermally anchored to the vacuum chamber. We model the trap temperature distribution with our finite element method (FEM) model \cite{Dolezal2015}, allowing for precise determination of the BBR environment at the ion positions. The model is refined for the new AlN traps based on measurements of the thermal distribution using an infrared camera \cite{Didier2018}. At $\Omega_\mathrm{rf}=2\pi\times\unit[24.4]{MHz}$, the radial confinement of $\nu_\mathrm{rad,In}=\unit[750]{kHz}$ ($\unit[1.5]{MHz}$) for configuration \emph{A} (\emph{B}) requires an rf voltage amplitude of $\unit[750]{V}$ ($\unit[1.5]{kV}$). The resulting mean temperature increase inferred from thermistor measurements is $\unit[0.6]{K}$ ($\unit[2.2]{K}$), corresponding to an effective temperature increase of $\unit[0.2]{K}$ ($\unit[0.8]{K}$) at the ion positions. Its overall uncertainty of $\unit[0.08]{K}$ (\unit[0.3]{K}) gives rise to a $1.5\times10^{-20}$ ($5.4\times10^{-20}$) fractional frequency uncertainty.
To avoid an increased uncertainty due to contributions from the vacuum chamber, its temperature also needs to be stabilized at this level. Temperature control beyond these requirements has already been demonstrated, e.g.~in neutral atom clocks using an external BBR shield~\cite{Ludlow2015}.

Presently, the uncertainty in the theoretical value of the differential static polarizability $\Delta\alpha_\mathrm{stat}=\unitfrac[3.3(3)\times10^{-41}]{Jm^2}{V^2}$ contributes an overall uncertainty of $1\times10^{-18}$ \cite{Safronova2011} at $T=\unit[300]{K}$, making the BBR shift dominate the uncertainty budget. In the future, this could be reduced to below $1\times10^{-20}$ by measurements of IR laser-induced light shifts: applying $\unit[30]{mW}$ of laser power in a $\unit[450]{\mu m}$ beam waist corresponds to $\langle E^2\rangle\approx(\unitfrac[8.4]{kV}{m})^2$, inducing a fractional frequency shift of ~$1.4\times10^{-15}$. By resolving this shift to within $1\times10^{-18}$, the fractional uncertainty in $\Delta\alpha_\mathrm{stat}$ could be reduced to $7\times10^{-4}$, corresponding to a contribution of $1\times10^{-20}$ to the room-temperature BBR AC Stark shift uncertainty.

\subsection{Collision shift}
Background gas collisions can shift electronic energy levels and thereby introduce a phase shift in the spectroscopy signal. For a known composition of the gas, the shift can be derived from the respective molecular potentials \cite{Vutha2017} or determined experimentally by varying the partial pressure \cite{Jau2015}. To obtain a conservative estimate, we adapt the treatment of \cite{Rosenband2008} to Ramsey spectroscopy. From Eq.~(\ref{ramseysignal}) with $\Omega_1=\Omega_2=\Omega$, we obtain
\begin{align}
\label{ramseysignalsymmetric}
S&=\sin^2(\Omega t)\left[\cos^2\left(\frac{\Delta T_\mathrm{int}}{2}-\frac{\pi}{4}\right)-\cos^2\left(\frac{\Delta T_\mathrm{int}}{2}+\frac{\pi}{4}\right)\right]\\\nonumber
&=\sin^2(\Omega t)\sin(\Delta T_\mathrm{int})\;\textnormal{.}
\end{align}
On resonance, both terms in square brackets take the value $1/2$, such that the error signal vanishes. The most detrimental effect of a collision is thus a phase shift of $\pm\pi/4$, making either of the terms $0$ or $1$ (we neglect the unlikely possibility that two consecutive interrogations are affected by collisions in opposite ways). The erroneous frequency correction derived from such a signal is therefore
\begin{equation}
\delta\nu=\frac{\Delta}{2\pi}=\frac{\arcsin\left(\frac{1}{2}\right)}{2\pi T_\mathrm{int}}\approx\frac{0.083}{T_\mathrm{int}}\;\textnormal{.}
\end{equation}
An upper bound for the overall frequency shift can be obtained by multiplying this step with the probability of a collision to occur in one of the two interrogations in (\ref{ramseysignalsymmetric}). Collisions involving more than one ion are likely to result in a detectable change of the fluorescence during Doppler cooling and state detection, such that data from the affected crystal can be disregarded in the respective clock cycle. In our systematic uncertainty estimate, we therefore assume that only one out of $N$ ions is affected by each collision. The overall collision shift is $\delta\nu/N$ times the probability of a collision to occur within two consecutive interrogation cycles:
\begin{align}
\Delta\nu_\mathrm{coll}=\frac{\delta\nu}{N}\frac{2T_\mathrm{int}}{T_\mathrm{coll}}\;\textnormal{,}
\end{align}
where $T_\mathrm{coll}$ is the average time between collisions. Since $T_\mathrm{coll}$ scales inversely with $N$ and $\delta\nu$ inversely with $T_{\mathrm{int}}$, $\Delta\nu_\mathrm{coll}$ is independent of both the ion number and interrogation time under our assumptions. We experimentally determine the average collision rate per ion by observing changes in the crystal order of a two-ion \ion{In}/\ion{Yb} crystal over $\unit[44]{h}$. With the derived collision rate of $\unit[4.8\times10^{-3}]{s^{-1}}$ per ion, the above estimate yields a worst-case fractional frequency shift of $6.3\times10^{-19}$. This simple estimate shows that while the collision shift requires further investigation to prevent it from dominating the overall clock uncertainty, reducing its uncertainty to below $1\times10^{-19}$ seems within reach.

\subsection{\label{magnetic_field}Magnetic field inhomogeneities}
Clock spectroscopy on a spatially extended ion ensemble requires good control of the magnetic field homogeneity over the trapping region; position-dependent resonance frequencies will otherwise result in inhomogeneous broadening and systematic shifts. The \ionm{115}{In} isotope has a nuclear spin of $I=9/2$. Starting clock interrogation from either stretched state of the ground state (${}^1$S${}_0$, $\vert m_F\vert=9/2$) to simplify state preparation, the minimal Zeeman shift occurs in the transition to the respective ${}^3$P${}_0$, $\vert m_F\vert=7/2$ state. All following estimates refer to this transition, the first-order Zeeman sensitivity of which is $\unitfrac[6.5]{mHz}{nT}$, about three orders of magnitude smaller than for atomic states with an electronic angular momentum $J>0$.

To ensure first-order Zeeman shift deviations below the natural linewidth of $\unit[820]{mHz}$ across the trapping region, we aim at field variations of less than $\unit[120]{nT}$ over $\unit[7]{mm}$ in our setup. With a 2-layer magnetic shield and permanent magnets, gradients of $\unit[250]{nT}$ along a trapping region of $\unit[6]{mm}$ have been observed using a \ionm{40}{Ca}-ion as a field probe \cite{Ruster2016,Ruster2017}. Considering the manufacturing tolerances of our Helmholtz-like coils, the gradient for an applied field of $\unit[100]{\mu T}$ is expected to be $\unit[1]{nT}$ ($\unit[13]{nT}$) across a $\unit[400]{\mu m}$ crystal (the whole $\unit[7]{mm}$ trapping region), corresponding to a first-order Zeeman spread of $\unit[7]{mHz}$ ($\unit[85]{mHz}$). Static spatial inhomogeneities will be mapped out using \ion{Yb} ion probes and can be accounted for in the frequency corrections applied by the servo algorithm. Commercially available power supplies for the coils achieve a relative RMS current noise of $4\times10^{-5}$, translating to a broadening of $\unit[26]{mHz}$, which is sufficiently below our resolution.

The second-order Zeeman shift is given by $\Delta\nu_\mathrm{Z2}=\beta\langle B^2\rangle$, where for \ionm{115}{In}, $\beta=2\mu^2_B/(3h^2\Delta_\mathrm{FS})=\unitfrac[4.1]{Hz}{mT^2}$. As a consequence of clock states without hyperfine structure and a comparably large fine-structure splitting $\Delta_\mathrm{FS}$ of the $^3$P manifold, the clock transition has a favorably low second-order magnetic field sensitivity, e.g.~3 to 4 orders of magnitude lower than those in \ionm{171}{Yb} \cite{Godun2014}. In the case of unbalanced trap rf currents, which have been observed to produce alternating magnetic fields on the order of $B^2_\mathrm{rms}=\unit[2.2\times 10^{-11}]{T^2}$ \cite{Chou2010a}, the resulting fractional frequency shift would be $7\times 10^{-20}$. 

\section{\label{conclusion_section}Conclusion}

\begingroup
\squeezetable
\begin{table*}
  \begin{ruledtabular}
      \begin{tabular}{@{}lrrrp{.01\textwidth}rrrr@{}} 
      &\multicolumn{3}{c}{narrow line cooled (\emph{A})}&&\multicolumn{3}{c}{sympathetically cooled (\emph{B})} \\ \cmidrule(r){2-4}\cmidrule(r){6-8}
      \textbf{values in $\mathbf{10^{-19}}$}&max. shift&spread&uncertainty&&max. shift&spread&uncertainty&comment\tabularnewline\toprule
      Time dilation (thermal)\footnote{See \cite{Supplemental} for a derivation based on the mass defect for a single ion}&$-2$&$<0.1$&$0.4$&&$-19$&$5$&$4$&assuming $\sigma(T)/T=\unit[20]{\%}$\\\midrule
      Heating (per second)&$-3.1$&&$0.2$&&$-0.6$&&$0.02$&\\\midrule
      Time dilation (EMM)&$-1.8$&$<1$&$0.8$&&$-1.3$&$<0.5$&$0.6$&\\\midrule
      AC Stark (thermal MM)&$-0.003$&&&&$-0.03$&&&\\\midrule
      AC Stark (EMM)&$-0.2$&$0.1$&$0.1$&&$-0.2$&$0.1$&$0.1$&\\\midrule
      El. quadrupole shift&$-0.02$&$0.02$&$<0.01$&&$-1.1$&$0.9$&$0.02$&without $\sigma_\mathrm{theo}(\Theta({}^3$P${}_0))$ \cite{Beloy2017}\\\midrule
      BBR at $\unit[300]{K}$&$-137$&&$0.15$&&$-137$&&$0.54$&without $\sigma_\mathrm{theo}(\Delta\alpha_\mathrm{stat})$\footnote{The contribution from $\Delta\alpha_\mathrm{stat}$ can be experimentally reduced to below $10^{-20}$ (see section \ref{BBR_section})}\\\bottomrule\\
      Sum &$-141.3$&&$\mathbf{0.9}$&&$-158.7$&&$\mathbf{4.1}$&
    \end{tabular}
  \end{ruledtabular}
\caption{\label{uncertaintybudget}Multi-ion operation related uncertainty contributions for linear ion chains with 10 \ionm{115}{In} clock ions each, as shown in Fig.~\ref{coulombcrystals}b. The trap parameters are $\Omega_\mathrm{rf}=2\pi\times\unit[24.4]{MHz}$, $\nu_\mathrm{rad,In}=\unit[0.75]{MHz}$ ($\unit[1.5]{MHz}$) and $\nu_\mathrm{ax,In}=\unit[30]{kHz}$ ($\unit[205]{kHz}$), respectively. All values are given as fractional frequency deviations in units of $1\times10^{-19}$. ``Max. shift'' corresponds to the most affected ion, ``spread'' denotes the difference between the most and least affected ion.}
\end{table*}
\endgroup

In summary, we have presented a scalable ion trap array designed for simultaneous precision spectroscopy on separately trapped ion Coulomb crystals. Particular emphasis was placed on minimizing axial rf electric fields through a high level of symmetry and intrinsically low manufacturing tolerances. Material choices and modeling of the thermal environment enable strong rf confinement in the Lamb-Dicke regime, while keeping the BBR-induced AC Stark shift uncertainty due to the associated heating of the trap below $1\times10^{-19}$.

We consider the use of linear Coulomb crystals with mixed ion species as high-precision frequency references. Based on experimental results in the new rf trap array, we have derived expected frequency uncertainties in an \ion{In} multi-ion clock with \ion{Yb} cooling ions. The results are summarized in Table \ref{uncertaintybudget}, both for a configuration employing cooling on the narrow ${}^1\mathrm{S}_0\;\leftrightarrow\;{}^3\mathrm{P}_1$ line in \ion{In} (\emph{A}) and one relying solely on sympathetic cooling with \ion{Yb} (\emph{B}). As introduced in section \ref{experimental_setup}, these cooling schemes operate either in weak (\emph{A}) or strong (\emph{B}) Coulomb coupling regimes. Several systematic shifts benefit from weak coupling, i.e.~a small ratio of axial to radial confinement: a larger ion spacing reduces electric field gradients within the chain and the resulting electric quadrupole shifts. The lower influence of mutual repulsion on the radial potential also reduces the downward shift of higher-order radial mode frequencies, which results in lower time dilation and AC Stark shifts due to intrinsic micromotion at a given temperature. If the interrogation laser has a radial component, higher radial mode frequencies also reduce the influence of the Debye-Waller effect on clock instability. On the other hand, weak axial confinement can make the axial modes more prone to external heating: while the time dilation shift only depends on the increase in energy and is thus in principle independent of the mode frequency, the underlying electric field noise spectra typically have an inverse frequency dependence. Weak Coulomb coupling also reduces the effectiveness of sympathetic cooling. Finally, weak axial confinement leads to increased chain lengths, and therefore requires longer regions with small perturbations. Our results are based on constant trap parameters which take all these influences into account. Further optimization could be achieved with the use of separate cooling and interrogation configurations.

The results presented in this work will pave the way for multi-ion clock operation with a lower fundamental limit to statistical uncertainty than the current generation of single-ion clocks and the potential to implement novel clock schemes which promise further stability improvements. We show that such clocks could operate with relative systematic uncertainties at the $10^{-19}$ level using separate strings with ca.~10 clock ions. Scaling of the trap array to support independent confinement of 10 such chains, i.e.~100 clock ions, is straightforward. Our concept is suitable for all clock transitions featuring a small differential electric quadrupole moment, allowing multi-ion clocks based on, e.g.~\ion{Al} \cite{Schulte2015} or \ionm{229}{Th} \cite{Peik2003}, and other species with insensitive transitions that are designed via collective states \cite{Roos2006}. In the absence of a narrow cooling line, advanced cooling methods \cite{Lechner2016,Ejtemaee2017,Scharnhorst2017} can be employed. The ${}^2$F${}_{7/2}$ state in \ion{Yb} features a quadrupole moment that is small enough to reach relative shifts below $5\times10^{-17}$ \cite{Huntemann2012} for chains with 10 ions confined at $\nu_\mathrm{ax}=\unit[20]{kHz}$, also allowing an electric quadrupole shift uncertainty smaller than $10^{-18}$. The increased sensitivity to variations of clock laser intensity will require spatial beam shaping and could be mitigated with advanced interrogation protocols \cite{Hobson2016,Sanner2018,Yudin2018}.

\begin{acknowledgments}
The authors thank the PTB departments 5.3 and 5.5 for the collaboration on trap fabrication. This work was supported by DFG through grant ME3648/1-1 and SFB 1227 (DQ-mat), project B03.
\end{acknowledgments}
\clearpage
\bibliography{cc_clock_papers}

%merlin.mbs apsrev4-1.bst 2010-07-25 4.21a (PWD, AO, DPC) hacked
%Control: key (0)
%Control: author (8) initials jnrlst
%Control: editor formatted (1) identically to author
%Control: production of article title (-1) disabled
%Control: page (0) single
%Control: year (1) truncated
%Control: production of eprint (0) enabled
\begin{thebibliography}{79}%
\makeatletter
\providecommand \@ifxundefined [1]{%
 \@ifx{#1\undefined}
}%
\providecommand \@ifnum [1]{%
 \ifnum #1\expandafter \@firstoftwo
 \else \expandafter \@secondoftwo
 \fi
}%
\providecommand \@ifx [1]{%
 \ifx #1\expandafter \@firstoftwo
 \else \expandafter \@secondoftwo
 \fi
}%
\providecommand \natexlab [1]{#1}%
\providecommand \enquote  [1]{``#1''}%
\providecommand \bibnamefont  [1]{#1}%
\providecommand \bibfnamefont [1]{#1}%
\providecommand \citenamefont [1]{#1}%
\providecommand \href@noop [0]{\@secondoftwo}%
\providecommand \href [0]{\begingroup \@sanitize@url \@href}%
\providecommand \@href[1]{\@@startlink{#1}\@@href}%
\providecommand \@@href[1]{\endgroup#1\@@endlink}%
\providecommand \@sanitize@url [0]{\catcode `\\12\catcode `\$12\catcode
  `\&12\catcode `\#12\catcode `\^12\catcode `\_12\catcode `\%12\relax}%
\providecommand \@@startlink[1]{}%
\providecommand \@@endlink[0]{}%
\providecommand \url  [0]{\begingroup\@sanitize@url \@url }%
\providecommand \@url [1]{\endgroup\@href {#1}{\urlprefix }}%
\providecommand \urlprefix  [0]{URL }%
\providecommand \Eprint [0]{\href }%
\providecommand \doibase [0]{http://dx.doi.org/}%
\providecommand \selectlanguage [0]{\@gobble}%
\providecommand \bibinfo  [0]{\@secondoftwo}%
\providecommand \bibfield  [0]{\@secondoftwo}%
\providecommand \translation [1]{[#1]}%
\providecommand \BibitemOpen [0]{}%
\providecommand \bibitemStop [0]{}%
\providecommand \bibitemNoStop [0]{.\EOS\space}%
\providecommand \EOS [0]{\spacefactor3000\relax}%
\providecommand \BibitemShut  [1]{\csname bibitem#1\endcsname}%
\let\auto@bib@innerbib\@empty
%</preamble>
\bibitem [{\citenamefont {{Chou}}\ \emph {et~al.}(2010)\citenamefont {{Chou}},
  \citenamefont {{Hume}}, \citenamefont {{Koelemeij}}, \citenamefont
  {{Wineland}},\ and\ \citenamefont {{Rosenband}}}]{Chou2010a}%
  \BibitemOpen
  \bibfield  {author} {\bibinfo {author} {\bibfnamefont {C.~W.}\ \bibnamefont
  {{Chou}}}, \bibinfo {author} {\bibfnamefont {D.~B.}\ \bibnamefont {{Hume}}},
  \bibinfo {author} {\bibfnamefont {J.~C.~J.}\ \bibnamefont {{Koelemeij}}},
  \bibinfo {author} {\bibfnamefont {D.~J.}\ \bibnamefont {{Wineland}}}, \ and\
  \bibinfo {author} {\bibfnamefont {T.}~\bibnamefont {{Rosenband}}},\ }\href
  {\doibase 10.1103/PhysRevLett.104.070802} {\bibfield  {journal} {\bibinfo
  {journal} {Phys. Rev. Lett.}\ }\textbf {\bibinfo {volume} {104}},\ \bibinfo
  {pages} {070802} (\bibinfo {year} {2010})}\BibitemShut {NoStop}%
\bibitem [{\citenamefont {{Dub{\'e}}}\ \emph {et~al.}(2014)\citenamefont
  {{Dub{\'e}}}, \citenamefont {{Madej}}, \citenamefont {{Tibbo}},\ and\
  \citenamefont {{Bernard}}}]{Dube2014}%
  \BibitemOpen
  \bibfield  {author} {\bibinfo {author} {\bibfnamefont {P.}~\bibnamefont
  {{Dub{\'e}}}}, \bibinfo {author} {\bibfnamefont {A.~A.}\ \bibnamefont
  {{Madej}}}, \bibinfo {author} {\bibfnamefont {M.}~\bibnamefont {{Tibbo}}}, \
  and\ \bibinfo {author} {\bibfnamefont {J.~E.}\ \bibnamefont {{Bernard}}},\
  }\href {\doibase 10.1103/PhysRevLett.112.173002} {\bibfield  {journal}
  {\bibinfo  {journal} {Phys. Rev. Lett.}\ }\textbf {\bibinfo {volume} {112}},\
  \bibinfo {eid} {173002} (\bibinfo {year} {2014})}\BibitemShut {NoStop}%
\bibitem [{\citenamefont {{Huntemann}}\ \emph {et~al.}(2016)\citenamefont
  {{Huntemann}}, \citenamefont {{Sanner}}, \citenamefont {{Lipphardt}},
  \citenamefont {{Tamm}},\ and\ \citenamefont {{Peik}}}]{Huntemann2016}%
  \BibitemOpen
  \bibfield  {author} {\bibinfo {author} {\bibfnamefont {N.}~\bibnamefont
  {{Huntemann}}}, \bibinfo {author} {\bibfnamefont {C.}~\bibnamefont
  {{Sanner}}}, \bibinfo {author} {\bibfnamefont {B.}~\bibnamefont
  {{Lipphardt}}}, \bibinfo {author} {\bibfnamefont {C.}~\bibnamefont {{Tamm}}},
  \ and\ \bibinfo {author} {\bibfnamefont {E.}~\bibnamefont {{Peik}}},\ }\href
  {\doibase 10.1103/PhysRevLett.116.063001} {\bibfield  {journal} {\bibinfo
  {journal} {Phys. Rev. Lett.}\ }\textbf {\bibinfo {volume} {116}},\ \bibinfo
  {eid} {063001} (\bibinfo {year} {2016})}\BibitemShut {NoStop}%
\bibitem [{\citenamefont {{Ludlow}}\ \emph {et~al.}(2015)\citenamefont
  {{Ludlow}}, \citenamefont {{Boyd}}, \citenamefont {{Ye}}, \citenamefont
  {{Peik}},\ and\ \citenamefont {{Schmidt}}}]{Ludlow2014}%
  \BibitemOpen
  \bibfield  {author} {\bibinfo {author} {\bibfnamefont {A.~D.}\ \bibnamefont
  {{Ludlow}}}, \bibinfo {author} {\bibfnamefont {M.~M.}\ \bibnamefont
  {{Boyd}}}, \bibinfo {author} {\bibfnamefont {J.}~\bibnamefont {{Ye}}},
  \bibinfo {author} {\bibfnamefont {E.}~\bibnamefont {{Peik}}}, \ and\ \bibinfo
  {author} {\bibfnamefont {P.~O.}\ \bibnamefont {{Schmidt}}},\ }\href {\doibase
  10.1103/RevModPhys.87.637} {\bibfield  {journal} {\bibinfo  {journal} {Rev.
  Mod. Phys}\ }\textbf {\bibinfo {volume} {87}},\ \bibinfo {pages} {637}
  (\bibinfo {year} {2015})}\BibitemShut {NoStop}%
\bibitem [{\citenamefont {{Huntemann}}\ \emph {et~al.}(2014)\citenamefont
  {{Huntemann}}, \citenamefont {{Lipphardt}}, \citenamefont {{Tamm}},
  \citenamefont {{Gerginov}}, \citenamefont {{Weyers}},\ and\ \citenamefont
  {{Peik}}}]{Huntemann2014a}%
  \BibitemOpen
  \bibfield  {author} {\bibinfo {author} {\bibfnamefont {N.}~\bibnamefont
  {{Huntemann}}}, \bibinfo {author} {\bibfnamefont {B.}~\bibnamefont
  {{Lipphardt}}}, \bibinfo {author} {\bibfnamefont {C.}~\bibnamefont {{Tamm}}},
  \bibinfo {author} {\bibfnamefont {V.}~\bibnamefont {{Gerginov}}}, \bibinfo
  {author} {\bibfnamefont {S.}~\bibnamefont {{Weyers}}}, \ and\ \bibinfo
  {author} {\bibfnamefont {E.}~\bibnamefont {{Peik}}},\ }\href {\doibase
  10.1103/PhysRevLett.113.210802} {\bibfield  {journal} {\bibinfo  {journal}
  {Phys. Rev. Lett.}\ }\textbf {\bibinfo {volume} {113}},\ \bibinfo {eid}
  {210802} (\bibinfo {year} {2014})}\BibitemShut {NoStop}%
\bibitem [{\citenamefont {{Godun}}\ \emph {et~al.}(2014)\citenamefont
  {{Godun}}, \citenamefont {{Nisbet-Jones}}, \citenamefont {{Jones}},
  \citenamefont {{King}}, \citenamefont {{Johnson}}, \citenamefont
  {{Margolis}}, \citenamefont {{Szymaniec}}, \citenamefont {{Lea}},
  \citenamefont {{Bongs}},\ and\ \citenamefont {{Gill}}}]{Godun2014}%
  \BibitemOpen
  \bibfield  {author} {\bibinfo {author} {\bibfnamefont {R.~M.}\ \bibnamefont
  {{Godun}}}, \bibinfo {author} {\bibfnamefont {P.~B.~R.}\ \bibnamefont
  {{Nisbet-Jones}}}, \bibinfo {author} {\bibfnamefont {J.~M.}\ \bibnamefont
  {{Jones}}}, \bibinfo {author} {\bibfnamefont {S.~A.}\ \bibnamefont {{King}}},
  \bibinfo {author} {\bibfnamefont {L.~A.~M.}\ \bibnamefont {{Johnson}}},
  \bibinfo {author} {\bibfnamefont {H.~S.}\ \bibnamefont {{Margolis}}},
  \bibinfo {author} {\bibfnamefont {K.}~\bibnamefont {{Szymaniec}}}, \bibinfo
  {author} {\bibfnamefont {S.~N.}\ \bibnamefont {{Lea}}}, \bibinfo {author}
  {\bibfnamefont {K.}~\bibnamefont {{Bongs}}}, \ and\ \bibinfo {author}
  {\bibfnamefont {P.}~\bibnamefont {{Gill}}},\ }\href {\doibase
  10.1103/PhysRevLett.113.210801} {\bibfield  {journal} {\bibinfo  {journal}
  {Phys. Rev. Lett.}\ }\textbf {\bibinfo {volume} {113}},\ \bibinfo {eid}
  {210801} (\bibinfo {year} {2014})}\BibitemShut {NoStop}%
\bibitem [{\citenamefont {{Pruttivarasin}}\ \emph {et~al.}(2015)\citenamefont
  {{Pruttivarasin}}, \citenamefont {{Ramm}}, \citenamefont {{Porsev}},
  \citenamefont {{Tupitsyn}}, \citenamefont {{Safronova}}, \citenamefont
  {{Hohensee}},\ and\ \citenamefont {{H{\"a}ffner}}}]{Pruttivarasin2015}%
  \BibitemOpen
  \bibfield  {author} {\bibinfo {author} {\bibfnamefont {T.}~\bibnamefont
  {{Pruttivarasin}}}, \bibinfo {author} {\bibfnamefont {M.}~\bibnamefont
  {{Ramm}}}, \bibinfo {author} {\bibfnamefont {S.~G.}\ \bibnamefont
  {{Porsev}}}, \bibinfo {author} {\bibfnamefont {I.~I.}\ \bibnamefont
  {{Tupitsyn}}}, \bibinfo {author} {\bibfnamefont {M.~S.}\ \bibnamefont
  {{Safronova}}}, \bibinfo {author} {\bibfnamefont {M.~A.}\ \bibnamefont
  {{Hohensee}}}, \ and\ \bibinfo {author} {\bibfnamefont {H.}~\bibnamefont
  {{H{\"a}ffner}}},\ }\href {\doibase 10.1038/nature14091} {\bibfield
  {journal} {\bibinfo  {journal} {Nature}\ }\textbf {\bibinfo {volume} {517}},\
  \bibinfo {pages} {592} (\bibinfo {year} {2015})}\BibitemShut {NoStop}%
\bibitem [{\citenamefont {{Dzuba}}\ \emph {et~al.}(2016)\citenamefont
  {{Dzuba}}, \citenamefont {{Flambaum}}, \citenamefont {{Safronova}},
  \citenamefont {{Porsev}}, \citenamefont {{Pruttivarasin}}, \citenamefont
  {{Hohensee}},\ and\ \citenamefont {{H{\"a}ffner}}}]{Dzuba2016}%
  \BibitemOpen
  \bibfield  {author} {\bibinfo {author} {\bibfnamefont {V.~A.}\ \bibnamefont
  {{Dzuba}}}, \bibinfo {author} {\bibfnamefont {V.~V.}\ \bibnamefont
  {{Flambaum}}}, \bibinfo {author} {\bibfnamefont {M.~S.}\ \bibnamefont
  {{Safronova}}}, \bibinfo {author} {\bibfnamefont {S.~G.}\ \bibnamefont
  {{Porsev}}}, \bibinfo {author} {\bibfnamefont {T.}~\bibnamefont
  {{Pruttivarasin}}}, \bibinfo {author} {\bibfnamefont {M.~A.}\ \bibnamefont
  {{Hohensee}}}, \ and\ \bibinfo {author} {\bibfnamefont {H.}~\bibnamefont
  {{H{\"a}ffner}}},\ }\href {\doibase 10.1038/nphys3610} {\bibfield  {journal}
  {\bibinfo  {journal} {Nat. Phys.}\ }\textbf {\bibinfo {volume} {12}},\
  \bibinfo {pages} {465} (\bibinfo {year} {2016})}\BibitemShut {NoStop}%
\bibitem [{\citenamefont {{Delva}}\ \emph {et~al.}(2017)\citenamefont
  {{Delva}}, \citenamefont {{Lodewyck}}, \citenamefont {{Bilicki}},
  \citenamefont {{Bookjans}}, \citenamefont {{Vallet}}, \citenamefont {{Le
  Targat}}, \citenamefont {{Pottie}}, \citenamefont {{Guerlin}}, \citenamefont
  {{Meynadier}}, \citenamefont {{Le Poncin-Lafitte}}, \citenamefont {{Lopez}},
  \citenamefont {{Amy-Klein}}, \citenamefont {{Lee}}, \citenamefont
  {{Quintin}}, \citenamefont {{Lisdat}}, \citenamefont {{Al-Masoudi}},
  \citenamefont {{D{\"o}rscher}}, \citenamefont {{Grebing}}, \citenamefont
  {{Grosche}}, \citenamefont {{Kuhl}}, \citenamefont {{Raupach}}, \citenamefont
  {{Sterr}}, \citenamefont {{Hill}}, \citenamefont {{Hobson}}, \citenamefont
  {{Bowden}}, \citenamefont {{Kronj{\"a}ger}}, \citenamefont {{Marra}},
  \citenamefont {{Rolland}}, \citenamefont {{Baynes}}, \citenamefont
  {{Margolis}},\ and\ \citenamefont {{Gill}}}]{Delva2017}%
  \BibitemOpen
  \bibfield  {author} {\bibinfo {author} {\bibfnamefont {P.}~\bibnamefont
  {{Delva}}}, \bibinfo {author} {\bibfnamefont {J.}~\bibnamefont {{Lodewyck}}},
  \bibinfo {author} {\bibfnamefont {S.}~\bibnamefont {{Bilicki}}}, \bibinfo
  {author} {\bibfnamefont {E.}~\bibnamefont {{Bookjans}}}, \bibinfo {author}
  {\bibfnamefont {G.}~\bibnamefont {{Vallet}}}, \bibinfo {author}
  {\bibfnamefont {R.}~\bibnamefont {{Le Targat}}}, \bibinfo {author}
  {\bibfnamefont {P.-E.}\ \bibnamefont {{Pottie}}}, \bibinfo {author}
  {\bibfnamefont {C.}~\bibnamefont {{Guerlin}}}, \bibinfo {author}
  {\bibfnamefont {F.}~\bibnamefont {{Meynadier}}}, \bibinfo {author}
  {\bibfnamefont {C.}~\bibnamefont {{Le Poncin-Lafitte}}}, \bibinfo {author}
  {\bibfnamefont {O.}~\bibnamefont {{Lopez}}}, \bibinfo {author} {\bibfnamefont
  {A.}~\bibnamefont {{Amy-Klein}}}, \bibinfo {author} {\bibfnamefont {W.-K.}\
  \bibnamefont {{Lee}}}, \bibinfo {author} {\bibfnamefont {N.}~\bibnamefont
  {{Quintin}}}, \bibinfo {author} {\bibfnamefont {C.}~\bibnamefont {{Lisdat}}},
  \bibinfo {author} {\bibfnamefont {A.}~\bibnamefont {{Al-Masoudi}}}, \bibinfo
  {author} {\bibfnamefont {S.}~\bibnamefont {{D{\"o}rscher}}}, \bibinfo
  {author} {\bibfnamefont {C.}~\bibnamefont {{Grebing}}}, \bibinfo {author}
  {\bibfnamefont {G.}~\bibnamefont {{Grosche}}}, \bibinfo {author}
  {\bibfnamefont {A.}~\bibnamefont {{Kuhl}}}, \bibinfo {author} {\bibfnamefont
  {S.}~\bibnamefont {{Raupach}}}, \bibinfo {author} {\bibfnamefont
  {U.}~\bibnamefont {{Sterr}}}, \bibinfo {author} {\bibfnamefont {I.~R.}\
  \bibnamefont {{Hill}}}, \bibinfo {author} {\bibfnamefont {R.}~\bibnamefont
  {{Hobson}}}, \bibinfo {author} {\bibfnamefont {W.}~\bibnamefont {{Bowden}}},
  \bibinfo {author} {\bibfnamefont {J.}~\bibnamefont {{Kronj{\"a}ger}}},
  \bibinfo {author} {\bibfnamefont {G.}~\bibnamefont {{Marra}}}, \bibinfo
  {author} {\bibfnamefont {A.}~\bibnamefont {{Rolland}}}, \bibinfo {author}
  {\bibfnamefont {F.~N.}\ \bibnamefont {{Baynes}}}, \bibinfo {author}
  {\bibfnamefont {H.~S.}\ \bibnamefont {{Margolis}}}, \ and\ \bibinfo {author}
  {\bibfnamefont {P.}~\bibnamefont {{Gill}}},\ }\href {\doibase
  10.1103/PhysRevLett.118.221102} {\bibfield  {journal} {\bibinfo  {journal}
  {Phys. Rev. Lett.}\ }\textbf {\bibinfo {volume} {118}},\ \bibinfo {eid}
  {221102} (\bibinfo {year} {2017})}\BibitemShut {NoStop}%
\bibitem [{\citenamefont {{Bjerhammar}}(1985)}]{Bjerhammar1985}%
  \BibitemOpen
  \bibfield  {author} {\bibinfo {author} {\bibfnamefont {A.}~\bibnamefont
  {{Bjerhammar}}},\ }\href {\doibase 10.1007/BF02520327} {\bibfield  {journal}
  {\bibinfo  {journal} {B. Geod.}\ }\textbf {\bibinfo {volume} {59}},\ \bibinfo
  {pages} {207} (\bibinfo {year} {1985})}\BibitemShut {NoStop}%
\bibitem [{\citenamefont {{Vermeer}}(1983)}]{Vermeer1983}%
  \BibitemOpen
  \bibfield  {author} {\bibinfo {author} {\bibfnamefont {M.}~\bibnamefont
  {{Vermeer}}},\ }\href@noop {} {\bibfield  {journal} {\bibinfo  {journal}
  {Report of the Finnish Geodetic Institute}\ }\textbf {\bibinfo {volume}
  {83}},\ \bibinfo {pages} {1} (\bibinfo {year} {1983})}\BibitemShut {NoStop}%
\bibitem [{\citenamefont {{Mehlst\"aubler}}\ \emph {et~al.}(2018)\citenamefont
  {{Mehlst\"aubler}}, \citenamefont {{Grosche}}, \citenamefont {{Lisdat}},
  \citenamefont {{Schmidt}},\ and\ \citenamefont
  {{Denker}}}]{Mehlstaeubler2017}%
  \BibitemOpen
  \bibfield  {author} {\bibinfo {author} {\bibfnamefont {T.~E.}\ \bibnamefont
  {{Mehlst\"aubler}}}, \bibinfo {author} {\bibfnamefont {G.}~\bibnamefont
  {{Grosche}}}, \bibinfo {author} {\bibfnamefont {C.}~\bibnamefont {{Lisdat}}},
  \bibinfo {author} {\bibfnamefont {P.~O.}\ \bibnamefont {{Schmidt}}}, \ and\
  \bibinfo {author} {\bibfnamefont {H.}~\bibnamefont {{Denker}}},\ }\href
  {\doibase 10.1088/1361-6633/aab409} {\bibfield  {journal} {\bibinfo
  {journal} {Rep. Prog. Phys.}\ }\textbf {\bibinfo {volume} {81}},\ \bibinfo
  {pages} {064401} (\bibinfo {year} {2018})}\BibitemShut {NoStop}%
\bibitem [{\citenamefont {{Itano}}\ \emph {et~al.}(1993)\citenamefont
  {{Itano}}, \citenamefont {{Bergquist}}, \citenamefont {{Bollinger}},
  \citenamefont {{Gilligan}}, \citenamefont {{Heinzen}}, \citenamefont
  {{Moore}}, \citenamefont {{Raizen}},\ and\ \citenamefont
  {{Wineland}}}]{Itano1993}%
  \BibitemOpen
  \bibfield  {author} {\bibinfo {author} {\bibfnamefont {W.~M.}\ \bibnamefont
  {{Itano}}}, \bibinfo {author} {\bibfnamefont {J.~C.}\ \bibnamefont
  {{Bergquist}}}, \bibinfo {author} {\bibfnamefont {J.~J.}\ \bibnamefont
  {{Bollinger}}}, \bibinfo {author} {\bibfnamefont {J.~M.}\ \bibnamefont
  {{Gilligan}}}, \bibinfo {author} {\bibfnamefont {D.~J.}\ \bibnamefont
  {{Heinzen}}}, \bibinfo {author} {\bibfnamefont {F.~L.}\ \bibnamefont
  {{Moore}}}, \bibinfo {author} {\bibfnamefont {M.~G.}\ \bibnamefont
  {{Raizen}}}, \ and\ \bibinfo {author} {\bibfnamefont {D.~J.}\ \bibnamefont
  {{Wineland}}},\ }\href {\doibase 10.1103/PhysRevA.47.3554} {\bibfield
  {journal} {\bibinfo  {journal} {Phys. Rev. A}\ }\textbf {\bibinfo {volume}
  {47}},\ \bibinfo {pages} {3554} (\bibinfo {year} {1993})}\BibitemShut
  {NoStop}%
\bibitem [{\citenamefont {{Champenois}}\ \emph {et~al.}(2010)\citenamefont
  {{Champenois}}, \citenamefont {{Marciante}}, \citenamefont
  {{Pedregosa-Gutierrez}}, \citenamefont {{Houssin}}, \citenamefont {{Knoop}},\
  and\ \citenamefont {{Kajita}}}]{Champenois2010}%
  \BibitemOpen
  \bibfield  {author} {\bibinfo {author} {\bibfnamefont {C.}~\bibnamefont
  {{Champenois}}}, \bibinfo {author} {\bibfnamefont {M.}~\bibnamefont
  {{Marciante}}}, \bibinfo {author} {\bibfnamefont {J.}~\bibnamefont
  {{Pedregosa-Gutierrez}}}, \bibinfo {author} {\bibfnamefont {M.}~\bibnamefont
  {{Houssin}}}, \bibinfo {author} {\bibfnamefont {M.}~\bibnamefont {{Knoop}}},
  \ and\ \bibinfo {author} {\bibfnamefont {M.}~\bibnamefont {{Kajita}}},\
  }\href {\doibase 10.1103/PhysRevA.81.043410} {\bibfield  {journal} {\bibinfo
  {journal} {Phys. Rev. A}\ }\textbf {\bibinfo {volume} {81}},\ \bibinfo {eid}
  {043410} (\bibinfo {year} {2010})}\BibitemShut {NoStop}%
\bibitem [{\citenamefont {{Herschbach}}\ \emph {et~al.}(2012)\citenamefont
  {{Herschbach}}, \citenamefont {{Pyka}}, \citenamefont {{Keller}},\ and\
  \citenamefont {{Mehlst{\"a}ubler}}}]{Herschbach2012}%
  \BibitemOpen
  \bibfield  {author} {\bibinfo {author} {\bibfnamefont {N.}~\bibnamefont
  {{Herschbach}}}, \bibinfo {author} {\bibfnamefont {K.}~\bibnamefont
  {{Pyka}}}, \bibinfo {author} {\bibfnamefont {J.}~\bibnamefont {{Keller}}}, \
  and\ \bibinfo {author} {\bibfnamefont {T.~E.}\ \bibnamefont
  {{Mehlst{\"a}ubler}}},\ }\href {\doibase 10.1007/s00340-011-4790-y}
  {\bibfield  {journal} {\bibinfo  {journal} {Appl. Phys. B}\ }\textbf
  {\bibinfo {volume} {107}},\ \bibinfo {pages} {891} (\bibinfo {year}
  {2012})}\BibitemShut {NoStop}%
\bibitem [{\citenamefont {{Arnold}}\ \emph {et~al.}(2015)\citenamefont
  {{Arnold}}, \citenamefont {{Hajiyev}}, \citenamefont {{Paez}}, \citenamefont
  {{Lee}}, \citenamefont {{Barrett}},\ and\ \citenamefont
  {{Bollinger}}}]{Arnold2015}%
  \BibitemOpen
  \bibfield  {author} {\bibinfo {author} {\bibfnamefont {K.}~\bibnamefont
  {{Arnold}}}, \bibinfo {author} {\bibfnamefont {E.}~\bibnamefont {{Hajiyev}}},
  \bibinfo {author} {\bibfnamefont {E.}~\bibnamefont {{Paez}}}, \bibinfo
  {author} {\bibfnamefont {C.~H.}\ \bibnamefont {{Lee}}}, \bibinfo {author}
  {\bibfnamefont {M.~D.}\ \bibnamefont {{Barrett}}}, \ and\ \bibinfo {author}
  {\bibfnamefont {J.}~\bibnamefont {{Bollinger}}},\ }\href {\doibase
  10.1103/PhysRevA.92.032108} {\bibfield  {journal} {\bibinfo  {journal} {Phys.
  Rev. A}\ }\textbf {\bibinfo {volume} {92}},\ \bibinfo {pages} {032108}
  (\bibinfo {year} {2015})}\BibitemShut {NoStop}%
\bibitem [{\citenamefont {{Rosenband}}\ and\ \citenamefont
  {{Leibrandt}}(2013)}]{Rosenband2013}%
  \BibitemOpen
  \bibfield  {author} {\bibinfo {author} {\bibfnamefont {T.}~\bibnamefont
  {{Rosenband}}}\ and\ \bibinfo {author} {\bibfnamefont {D.~R.}\ \bibnamefont
  {{Leibrandt}}},\ }\href@noop {} {\bibfield  {journal} {\bibinfo  {journal}
  {ArXiv e-prints}\ } (\bibinfo {year} {2013})},\ \Eprint
  {http://arxiv.org/abs/1303.6357} {arXiv:1303.6357} \BibitemShut {NoStop}%
\bibitem [{\citenamefont {{Borregaard}}\ and\ \citenamefont
  {{S{\o}rensen}}(2013)}]{Borregaard2013}%
  \BibitemOpen
  \bibfield  {author} {\bibinfo {author} {\bibfnamefont {J.}~\bibnamefont
  {{Borregaard}}}\ and\ \bibinfo {author} {\bibfnamefont {A.~S.}\ \bibnamefont
  {{S{\o}rensen}}},\ }\href {\doibase 10.1103/PhysRevLett.111.090802}
  {\bibfield  {journal} {\bibinfo  {journal} {Phys. Rev. Lett.}\ }\textbf
  {\bibinfo {volume} {111}},\ \bibinfo {eid} {090802} (\bibinfo {year}
  {2013})}\BibitemShut {NoStop}%
\bibitem [{\citenamefont {{Leroux}}\ \emph {et~al.}(2010)\citenamefont
  {{Leroux}}, \citenamefont {{Schleier-Smith}},\ and\ \citenamefont
  {{Vuleti{\'c}}}}]{Leroux2010}%
  \BibitemOpen
  \bibfield  {author} {\bibinfo {author} {\bibfnamefont {I.~D.}\ \bibnamefont
  {{Leroux}}}, \bibinfo {author} {\bibfnamefont {M.~H.}\ \bibnamefont
  {{Schleier-Smith}}}, \ and\ \bibinfo {author} {\bibfnamefont
  {V.}~\bibnamefont {{Vuleti{\'c}}}},\ }\href {\doibase
  10.1103/PhysRevLett.104.250801} {\bibfield  {journal} {\bibinfo  {journal}
  {Phys. Rev. Lett.}\ }\textbf {\bibinfo {volume} {104}},\ \bibinfo {pages}
  {250801} (\bibinfo {year} {2010})}\BibitemShut {NoStop}%
\bibitem [{\citenamefont {{Kessler}}\ \emph {et~al.}(2014)\citenamefont
  {{Kessler}}, \citenamefont {{K{\'o}m{\'a}r}}, \citenamefont {{Bishof}},
  \citenamefont {{Jiang}}, \citenamefont {{S{\o}rensen}}, \citenamefont
  {{Ye}},\ and\ \citenamefont {{Lukin}}}]{Kessler2014}%
  \BibitemOpen
  \bibfield  {author} {\bibinfo {author} {\bibfnamefont {E.~M.}\ \bibnamefont
  {{Kessler}}}, \bibinfo {author} {\bibfnamefont {P.}~\bibnamefont
  {{K{\'o}m{\'a}r}}}, \bibinfo {author} {\bibfnamefont {M.}~\bibnamefont
  {{Bishof}}}, \bibinfo {author} {\bibfnamefont {L.}~\bibnamefont {{Jiang}}},
  \bibinfo {author} {\bibfnamefont {A.~S.}\ \bibnamefont {{S{\o}rensen}}},
  \bibinfo {author} {\bibfnamefont {J.}~\bibnamefont {{Ye}}}, \ and\ \bibinfo
  {author} {\bibfnamefont {M.~D.}\ \bibnamefont {{Lukin}}},\ }\href {\doibase
  10.1103/PhysRevLett.112.190403} {\bibfield  {journal} {\bibinfo  {journal}
  {Phys. Rev. Lett.}\ }\textbf {\bibinfo {volume} {112}},\ \bibinfo {eid}
  {190403} (\bibinfo {year} {2014})}\BibitemShut {NoStop}%
\bibitem [{\citenamefont {{Lebedev}}\ \emph {et~al.}(2014)\citenamefont
  {{Lebedev}}, \citenamefont {{Treutlein}},\ and\ \citenamefont
  {{Blatter}}}]{Lebedev2014}%
  \BibitemOpen
  \bibfield  {author} {\bibinfo {author} {\bibfnamefont {A.~V.}\ \bibnamefont
  {{Lebedev}}}, \bibinfo {author} {\bibfnamefont {P.}~\bibnamefont
  {{Treutlein}}}, \ and\ \bibinfo {author} {\bibfnamefont {G.}~\bibnamefont
  {{Blatter}}},\ }\href {\doibase 10.1103/PhysRevA.89.012118} {\bibfield
  {journal} {\bibinfo  {journal} {Phys. Rev. A}\ }\textbf {\bibinfo {volume}
  {89}},\ \bibinfo {eid} {012118} (\bibinfo {year} {2014})}\BibitemShut
  {NoStop}%
\bibitem [{\citenamefont {{Schneider}}\ \emph {et~al.}(2012)\citenamefont
  {{Schneider}}, \citenamefont {{Porras}},\ and\ \citenamefont
  {{Schaetz}}}]{Schneider2012}%
  \BibitemOpen
  \bibfield  {author} {\bibinfo {author} {\bibfnamefont {C.}~\bibnamefont
  {{Schneider}}}, \bibinfo {author} {\bibfnamefont {D.}~\bibnamefont
  {{Porras}}}, \ and\ \bibinfo {author} {\bibfnamefont {T.}~\bibnamefont
  {{Schaetz}}},\ }\href {\doibase 10.1088/0034-4885/75/2/024401} {\bibfield
  {journal} {\bibinfo  {journal} {Rep. Prog. Phys.}\ }\textbf {\bibinfo
  {volume} {75}},\ \bibinfo {eid} {024401} (\bibinfo {year}
  {2012})}\BibitemShut {NoStop}%
\bibitem [{\citenamefont {{Blatt}}\ and\ \citenamefont
  {{Roos}}(2012)}]{Blatt2012}%
  \BibitemOpen
  \bibfield  {author} {\bibinfo {author} {\bibfnamefont {R.}~\bibnamefont
  {{Blatt}}}\ and\ \bibinfo {author} {\bibfnamefont {C.~F.}\ \bibnamefont
  {{Roos}}},\ }\href {\doibase 10.1038/nphys2252} {\bibfield  {journal}
  {\bibinfo  {journal} {Nat. Phys.}\ }\textbf {\bibinfo {volume} {8}},\
  \bibinfo {pages} {277} (\bibinfo {year} {2012})}\BibitemShut {NoStop}%
\bibitem [{\citenamefont {Hess}\ \emph {et~al.}(2017)\citenamefont {Hess},
  \citenamefont {Becker}, \citenamefont {Kaplan}, \citenamefont {Kyprianidis},
  \citenamefont {Lee}, \citenamefont {Neyenhuis}, \citenamefont {Pagano},
  \citenamefont {Richerme}, \citenamefont {Senko}, \citenamefont {Smith},
  \citenamefont {Tan}, \citenamefont {Zhang},\ and\ \citenamefont
  {Monroe}}]{Hess2017}%
  \BibitemOpen
  \bibfield  {author} {\bibinfo {author} {\bibfnamefont {P.~W.}\ \bibnamefont
  {Hess}}, \bibinfo {author} {\bibfnamefont {P.}~\bibnamefont {Becker}},
  \bibinfo {author} {\bibfnamefont {H.~B.}\ \bibnamefont {Kaplan}}, \bibinfo
  {author} {\bibfnamefont {A.}~\bibnamefont {Kyprianidis}}, \bibinfo {author}
  {\bibfnamefont {A.~C.}\ \bibnamefont {Lee}}, \bibinfo {author} {\bibfnamefont
  {B.}~\bibnamefont {Neyenhuis}}, \bibinfo {author} {\bibfnamefont
  {G.}~\bibnamefont {Pagano}}, \bibinfo {author} {\bibfnamefont
  {P.}~\bibnamefont {Richerme}}, \bibinfo {author} {\bibfnamefont
  {C.}~\bibnamefont {Senko}}, \bibinfo {author} {\bibfnamefont
  {J.}~\bibnamefont {Smith}}, \bibinfo {author} {\bibfnamefont {W.~L.}\
  \bibnamefont {Tan}}, \bibinfo {author} {\bibfnamefont {J.}~\bibnamefont
  {Zhang}}, \ and\ \bibinfo {author} {\bibfnamefont {C.}~\bibnamefont
  {Monroe}},\ }\href {\doibase 10.1098/rsta.2017.0107} {\bibfield  {journal}
  {\bibinfo  {journal} {Phil. Trans. R. Soc. A}\ }\textbf {\bibinfo {volume}
  {375}},\ \bibinfo {pages} {20170107} (\bibinfo {year} {2017})}\BibitemShut
  {NoStop}%
\bibitem [{\citenamefont {{Blatt}}\ and\ \citenamefont
  {{Wineland}}(2008)}]{Blatt2008}%
  \BibitemOpen
  \bibfield  {author} {\bibinfo {author} {\bibfnamefont {R.}~\bibnamefont
  {{Blatt}}}\ and\ \bibinfo {author} {\bibfnamefont {D.~J.}\ \bibnamefont
  {{Wineland}}},\ }\href {\doibase 10.1038/nature07125} {\bibfield  {journal}
  {\bibinfo  {journal} {Nature}\ }\textbf {\bibinfo {volume} {453}},\ \bibinfo
  {pages} {1008} (\bibinfo {year} {2008})}\BibitemShut {NoStop}%
\bibitem [{\citenamefont {{Becker}}\ \emph {et~al.}(2001)\citenamefont
  {{Becker}}, \citenamefont {{Zanthier}}, \citenamefont {{Nevsky}},
  \citenamefont {{Schwedes}}, \citenamefont {{Skvortsov}}, \citenamefont
  {{Walther}},\ and\ \citenamefont {{Peik}}}]{Becker2001}%
  \BibitemOpen
  \bibfield  {author} {\bibinfo {author} {\bibfnamefont {T.}~\bibnamefont
  {{Becker}}}, \bibinfo {author} {\bibfnamefont {J.~V.}\ \bibnamefont
  {{Zanthier}}}, \bibinfo {author} {\bibfnamefont {A.~Y.}\ \bibnamefont
  {{Nevsky}}}, \bibinfo {author} {\bibfnamefont {C.}~\bibnamefont
  {{Schwedes}}}, \bibinfo {author} {\bibfnamefont {M.~N.}\ \bibnamefont
  {{Skvortsov}}}, \bibinfo {author} {\bibfnamefont {H.}~\bibnamefont
  {{Walther}}}, \ and\ \bibinfo {author} {\bibfnamefont {E.}~\bibnamefont
  {{Peik}}},\ }\href {\doibase 10.1103/PhysRevA.63.051802} {\bibfield
  {journal} {\bibinfo  {journal} {Phys. Rev. A}\ }\textbf {\bibinfo {volume}
  {63}},\ \bibinfo {pages} {051802} (\bibinfo {year} {2001})}\BibitemShut
  {NoStop}%
\bibitem [{\citenamefont {{Ohtsubo}}\ \emph {et~al.}(2017)\citenamefont
  {{Ohtsubo}}, \citenamefont {{Li}}, \citenamefont {{Matsubara}}, \citenamefont
  {{Ido}},\ and\ \citenamefont {{Hayasaka}}}]{Ohtsubo2017}%
  \BibitemOpen
  \bibfield  {author} {\bibinfo {author} {\bibfnamefont {N.}~\bibnamefont
  {{Ohtsubo}}}, \bibinfo {author} {\bibfnamefont {Y.}~\bibnamefont {{Li}}},
  \bibinfo {author} {\bibfnamefont {K.}~\bibnamefont {{Matsubara}}}, \bibinfo
  {author} {\bibfnamefont {T.}~\bibnamefont {{Ido}}}, \ and\ \bibinfo {author}
  {\bibfnamefont {K.}~\bibnamefont {{Hayasaka}}},\ }\href {\doibase
  10.1364/OE.25.011725} {\bibfield  {journal} {\bibinfo  {journal} {Opt.
  Express}\ }\textbf {\bibinfo {volume} {25}},\ \bibinfo {pages} {11725}
  (\bibinfo {year} {2017})}\BibitemShut {NoStop}%
\bibitem [{\citenamefont {{Beloy}}\ \emph {et~al.}(2017)\citenamefont
  {{Beloy}}, \citenamefont {{Leibrandt}},\ and\ \citenamefont
  {{Itano}}}]{Beloy2017}%
  \BibitemOpen
  \bibfield  {author} {\bibinfo {author} {\bibfnamefont {K.}~\bibnamefont
  {{Beloy}}}, \bibinfo {author} {\bibfnamefont {D.~R.}\ \bibnamefont
  {{Leibrandt}}}, \ and\ \bibinfo {author} {\bibfnamefont {W.~M.}\ \bibnamefont
  {{Itano}}},\ }\href {\doibase 10.1103/PhysRevA.95.043405} {\bibfield
  {journal} {\bibinfo  {journal} {Phys. Rev. A}\ }\textbf {\bibinfo {volume}
  {95}},\ \bibinfo {pages} {043405} (\bibinfo {year} {2017})}\BibitemShut
  {NoStop}%
\bibitem [{\citenamefont {{Peik}}\ and\ \citenamefont
  {{Tamm}}(2003)}]{Peik2003}%
  \BibitemOpen
  \bibfield  {author} {\bibinfo {author} {\bibfnamefont {E.}~\bibnamefont
  {{Peik}}}\ and\ \bibinfo {author} {\bibfnamefont {C.}~\bibnamefont
  {{Tamm}}},\ }\href {\doibase 10.1209/epl/i2003-00210-x} {\bibfield  {journal}
  {\bibinfo  {journal} {EPL}\ }\textbf {\bibinfo {volume} {61}},\ \bibinfo
  {pages} {181} (\bibinfo {year} {2003})}\BibitemShut {NoStop}%
\bibitem [{\citenamefont {{Keller}}\ \emph {et~al.}(2017)\citenamefont
  {{Keller}}, \citenamefont {{Kalincev}}, \citenamefont {{Burgermeister}},
  \citenamefont {{Kulosa}}, \citenamefont {{Didier}}, \citenamefont
  {{Nordmann}}, \citenamefont {{Kiethe}},\ and\ \citenamefont
  {{Mehlst\"aubler}}}]{Keller2017}%
  \BibitemOpen
  \bibfield  {author} {\bibinfo {author} {\bibfnamefont {J.}~\bibnamefont
  {{Keller}}}, \bibinfo {author} {\bibfnamefont {D.}~\bibnamefont
  {{Kalincev}}}, \bibinfo {author} {\bibfnamefont {T.}~\bibnamefont
  {{Burgermeister}}}, \bibinfo {author} {\bibfnamefont {A.~P.}\ \bibnamefont
  {{Kulosa}}}, \bibinfo {author} {\bibfnamefont {A.}~\bibnamefont {{Didier}}},
  \bibinfo {author} {\bibfnamefont {T.}~\bibnamefont {{Nordmann}}}, \bibinfo
  {author} {\bibfnamefont {J.}~\bibnamefont {{Kiethe}}}, \ and\ \bibinfo
  {author} {\bibfnamefont {T.~E.}\ \bibnamefont {{Mehlst\"aubler}}},\
  }\href@noop {} {\bibfield  {journal} {\bibinfo  {journal} {ArXiv e-prints}\ }
  (\bibinfo {year} {2017})},\ \Eprint {http://arxiv.org/abs/1712.02335}
  {arXiv:1712.02335} \BibitemShut {NoStop}%
\bibitem [{\citenamefont {{Pyka}}\ \emph {et~al.}(2014)\citenamefont {{Pyka}},
  \citenamefont {{Herschbach}}, \citenamefont {{Keller}},\ and\ \citenamefont
  {{Mehlst{\"a}ubler}}}]{Pyka2014}%
  \BibitemOpen
  \bibfield  {author} {\bibinfo {author} {\bibfnamefont {K.}~\bibnamefont
  {{Pyka}}}, \bibinfo {author} {\bibfnamefont {N.}~\bibnamefont
  {{Herschbach}}}, \bibinfo {author} {\bibfnamefont {J.}~\bibnamefont
  {{Keller}}}, \ and\ \bibinfo {author} {\bibfnamefont {T.~E.}\ \bibnamefont
  {{Mehlst{\"a}ubler}}},\ }\href {\doibase 10.1007/s00340-013-5580-5}
  {\bibfield  {journal} {\bibinfo  {journal} {Appl. Phys. B}\ }\textbf
  {\bibinfo {volume} {114}},\ \bibinfo {pages} {231} (\bibinfo {year}
  {2014})}\BibitemShut {NoStop}%
\bibitem [{\citenamefont {{Kiethe}}\ \emph {et~al.}(2017)\citenamefont
  {{Kiethe}}, \citenamefont {{Nigmatullin}}, \citenamefont {{Kalincev}},
  \citenamefont {{Schmirander}},\ and\ \citenamefont
  {{Mehlst{\"a}ubler}}}]{Kiethe2017}%
  \BibitemOpen
  \bibfield  {author} {\bibinfo {author} {\bibfnamefont {J.}~\bibnamefont
  {{Kiethe}}}, \bibinfo {author} {\bibfnamefont {R.}~\bibnamefont
  {{Nigmatullin}}}, \bibinfo {author} {\bibfnamefont {D.}~\bibnamefont
  {{Kalincev}}}, \bibinfo {author} {\bibfnamefont {T.}~\bibnamefont
  {{Schmirander}}}, \ and\ \bibinfo {author} {\bibfnamefont {T.~E.}\
  \bibnamefont {{Mehlst{\"a}ubler}}},\ }\href {\doibase 10.1038/ncomms15364}
  {\bibfield  {journal} {\bibinfo  {journal} {Nat. Commun.}\ }\textbf {\bibinfo
  {volume} {8}},\ \bibinfo {eid} {15364} (\bibinfo {year} {2017})}\BibitemShut
  {NoStop}%
\bibitem [{\citenamefont {{Itano}}(2000)}]{Itano2000}%
  \BibitemOpen
  \bibfield  {author} {\bibinfo {author} {\bibfnamefont {W.~M.}\ \bibnamefont
  {{Itano}}},\ }\href {\doibase 10.6028/jres.105.065} {\bibfield  {journal}
  {\bibinfo  {journal} {J. Res. Natl. Inst. Stand. Technol.}\ }\textbf
  {\bibinfo {volume} {105}},\ \bibinfo {pages} {829} (\bibinfo {year}
  {2000})}\BibitemShut {NoStop}%
\bibitem [{\citenamefont {{Garstang}}(1962)}]{Garstang1962}%
  \BibitemOpen
  \bibfield  {author} {\bibinfo {author} {\bibfnamefont {R.~H.}\ \bibnamefont
  {{Garstang}}},\ }\href {\doibase 10.1364/JOSA.52.000845} {\bibfield
  {journal} {\bibinfo  {journal} {J. Opt. Soc. Am.}\ }\textbf {\bibinfo
  {volume} {52}},\ \bibinfo {pages} {845} (\bibinfo {year} {1962})}\BibitemShut
  {NoStop}%
\bibitem [{\citenamefont {{W{\"u}bbena}}\ \emph {et~al.}(2012)\citenamefont
  {{W{\"u}bbena}}, \citenamefont {{Amairi}}, \citenamefont {{Mandel}},\ and\
  \citenamefont {{Schmidt}}}]{Wuebbena2012}%
  \BibitemOpen
  \bibfield  {author} {\bibinfo {author} {\bibfnamefont {J.~B.}\ \bibnamefont
  {{W{\"u}bbena}}}, \bibinfo {author} {\bibfnamefont {S.}~\bibnamefont
  {{Amairi}}}, \bibinfo {author} {\bibfnamefont {O.}~\bibnamefont {{Mandel}}},
  \ and\ \bibinfo {author} {\bibfnamefont {P.~O.}\ \bibnamefont {{Schmidt}}},\
  }\href {\doibase 10.1103/PhysRevA.85.043412} {\bibfield  {journal} {\bibinfo
  {journal} {Phys. Rev. A}\ }\textbf {\bibinfo {volume} {85}},\ \bibinfo
  {pages} {043412} (\bibinfo {year} {2012})}\BibitemShut {NoStop}%
\bibitem [{\citenamefont {{James}}(1998)}]{James1998}%
  \BibitemOpen
  \bibfield  {author} {\bibinfo {author} {\bibfnamefont {D.~F.~V.}\
  \bibnamefont {{James}}},\ }\href {\doibase 10.1007/s003400050373} {\bibfield
  {journal} {\bibinfo  {journal} {Appl. Phys. B}\ }\textbf {\bibinfo {volume}
  {66}},\ \bibinfo {pages} {181} (\bibinfo {year} {1998})}\BibitemShut
  {NoStop}%
\bibitem [{\citenamefont {{Devanathan}}(2002)}]{Devanathan2002}%
  \BibitemOpen
  \bibfield  {author} {\bibinfo {author} {\bibfnamefont {V.}~\bibnamefont
  {{Devanathan}}},\ }\href@noop {} {\emph {\bibinfo {title} {{Angular momentum
  techniques in quantum mechanics}}}}\ (\bibinfo  {publisher} {Kluwer},\
  \bibinfo {year} {2002})\BibitemShut {NoStop}%
\bibitem [{\citenamefont {{Dub{\'e}}}\ \emph {et~al.}(2005)\citenamefont
  {{Dub{\'e}}}, \citenamefont {{Madej}}, \citenamefont {{Bernard}},
  \citenamefont {{Marmet}}, \citenamefont {{Boulanger}},\ and\ \citenamefont
  {{Cundy}}}]{Dube2005}%
  \BibitemOpen
  \bibfield  {author} {\bibinfo {author} {\bibfnamefont {P.}~\bibnamefont
  {{Dub{\'e}}}}, \bibinfo {author} {\bibfnamefont {A.~A.}\ \bibnamefont
  {{Madej}}}, \bibinfo {author} {\bibfnamefont {J.~E.}\ \bibnamefont
  {{Bernard}}}, \bibinfo {author} {\bibfnamefont {L.}~\bibnamefont {{Marmet}}},
  \bibinfo {author} {\bibfnamefont {J.-S.}\ \bibnamefont {{Boulanger}}}, \ and\
  \bibinfo {author} {\bibfnamefont {S.}~\bibnamefont {{Cundy}}},\ }\href
  {\doibase 10.1103/PhysRevLett.95.033001} {\bibfield  {journal} {\bibinfo
  {journal} {Phys. Rev. Lett.}\ }\textbf {\bibinfo {volume} {95}},\ \bibinfo
  {eid} {033001} (\bibinfo {year} {2005})}\BibitemShut {NoStop}%
\bibitem [{\citenamefont {{Barrett}}(2015)}]{Barrett2015}%
  \BibitemOpen
  \bibfield  {author} {\bibinfo {author} {\bibfnamefont {M.~D.}\ \bibnamefont
  {{Barrett}}},\ }\href {\doibase 10.1088/1367-2630/17/5/053024} {\bibfield
  {journal} {\bibinfo  {journal} {New. J. Phys.}\ }\textbf {\bibinfo {volume}
  {17}},\ \bibinfo {eid} {053024} (\bibinfo {year} {2015})}\BibitemShut
  {NoStop}%
\bibitem [{\citenamefont {{Berkeland}}\ \emph {et~al.}(1998)\citenamefont
  {{Berkeland}}, \citenamefont {{Miller}}, \citenamefont {{Bergquist}},
  \citenamefont {{Itano}},\ and\ \citenamefont {{Wineland}}}]{Berkeland1998}%
  \BibitemOpen
  \bibfield  {author} {\bibinfo {author} {\bibfnamefont {D.~J.}\ \bibnamefont
  {{Berkeland}}}, \bibinfo {author} {\bibfnamefont {J.~D.}\ \bibnamefont
  {{Miller}}}, \bibinfo {author} {\bibfnamefont {J.~C.}\ \bibnamefont
  {{Bergquist}}}, \bibinfo {author} {\bibfnamefont {W.~M.}\ \bibnamefont
  {{Itano}}}, \ and\ \bibinfo {author} {\bibfnamefont {D.~J.}\ \bibnamefont
  {{Wineland}}},\ }\href {\doibase 10.1063/1.367318} {\bibfield  {journal}
  {\bibinfo  {journal} {J. Appl. Phys.}\ }\textbf {\bibinfo {volume} {83}},\
  \bibinfo {pages} {5025} (\bibinfo {year} {1998})}\BibitemShut {NoStop}%
\bibitem [{\citenamefont {{Madej}}\ \emph {et~al.}(2012)\citenamefont
  {{Madej}}, \citenamefont {{Dub{\'e}}}, \citenamefont {{Zhou}}, \citenamefont
  {{Bernard}},\ and\ \citenamefont {{Gertsvolf}}}]{Madej2012}%
  \BibitemOpen
  \bibfield  {author} {\bibinfo {author} {\bibfnamefont {A.~A.}\ \bibnamefont
  {{Madej}}}, \bibinfo {author} {\bibfnamefont {P.}~\bibnamefont {{Dub{\'e}}}},
  \bibinfo {author} {\bibfnamefont {Z.}~\bibnamefont {{Zhou}}}, \bibinfo
  {author} {\bibfnamefont {J.~E.}\ \bibnamefont {{Bernard}}}, \ and\ \bibinfo
  {author} {\bibfnamefont {M.}~\bibnamefont {{Gertsvolf}}},\ }\href {\doibase
  10.1103/PhysRevLett.109.203002} {\bibfield  {journal} {\bibinfo  {journal}
  {Phys. Rev. Lett.}\ }\textbf {\bibinfo {volume} {109}},\ \bibinfo {pages}
  {203002} (\bibinfo {year} {2012})}\BibitemShut {NoStop}%
\bibitem [{\citenamefont {{Blakestad}}\ \emph {et~al.}(2009)\citenamefont
  {{Blakestad}}, \citenamefont {{Ospelkaus}}, \citenamefont {{VanDevender}},
  \citenamefont {{Amini}}, \citenamefont {{Britton}}, \citenamefont
  {{Leibfried}},\ and\ \citenamefont {{Wineland}}}]{Blakestad2009}%
  \BibitemOpen
  \bibfield  {author} {\bibinfo {author} {\bibfnamefont {R.~B.}\ \bibnamefont
  {{Blakestad}}}, \bibinfo {author} {\bibfnamefont {C.}~\bibnamefont
  {{Ospelkaus}}}, \bibinfo {author} {\bibfnamefont {A.~P.}\ \bibnamefont
  {{VanDevender}}}, \bibinfo {author} {\bibfnamefont {J.~M.}\ \bibnamefont
  {{Amini}}}, \bibinfo {author} {\bibfnamefont {J.}~\bibnamefont {{Britton}}},
  \bibinfo {author} {\bibfnamefont {D.}~\bibnamefont {{Leibfried}}}, \ and\
  \bibinfo {author} {\bibfnamefont {D.~J.}\ \bibnamefont {{Wineland}}},\ }\href
  {\doibase 10.1103/PhysRevLett.102.153002} {\bibfield  {journal} {\bibinfo
  {journal} {Phys. Rev. Lett.}\ }\textbf {\bibinfo {volume} {102}},\ \bibinfo
  {pages} {153002} (\bibinfo {year} {2009})}\BibitemShut {NoStop}%
\bibitem [{\citenamefont {{Morigi}}\ and\ \citenamefont
  {{Walther}}(2001)}]{Morigi2001}%
  \BibitemOpen
  \bibfield  {author} {\bibinfo {author} {\bibfnamefont {G.}~\bibnamefont
  {{Morigi}}}\ and\ \bibinfo {author} {\bibfnamefont {H.}~\bibnamefont
  {{Walther}}},\ }\href {\doibase 10.1007/s100530170275} {\bibfield  {journal}
  {\bibinfo  {journal} {Eur. Phys. J. D}\ }\textbf {\bibinfo {volume} {13}},\
  \bibinfo {pages} {261} (\bibinfo {year} {2001})}\BibitemShut {NoStop}%
\bibitem [{\citenamefont {{Home}}(2013)}]{Home2013}%
  \BibitemOpen
  \bibfield  {author} {\bibinfo {author} {\bibfnamefont {J.~P.}\ \bibnamefont
  {{Home}}},\ }\href {\doibase 10.1016/B978-0-12-408090-4.00004-9} {\bibfield
  {journal} {\bibinfo  {journal} {Adv. At. Mol. Opt. Phy.}\ }\textbf {\bibinfo
  {volume} {62}},\ \bibinfo {pages} {231} (\bibinfo {year} {2013})}\BibitemShut
  {NoStop}%
\bibitem [{\citenamefont {{Leibfried}}\ \emph {et~al.}(2003)\citenamefont
  {{Leibfried}}, \citenamefont {{Blatt}}, \citenamefont {{Monroe}},\ and\
  \citenamefont {{Wineland}}}]{Leibfried2003}%
  \BibitemOpen
  \bibfield  {author} {\bibinfo {author} {\bibfnamefont {D.}~\bibnamefont
  {{Leibfried}}}, \bibinfo {author} {\bibfnamefont {R.}~\bibnamefont
  {{Blatt}}}, \bibinfo {author} {\bibfnamefont {C.}~\bibnamefont {{Monroe}}}, \
  and\ \bibinfo {author} {\bibfnamefont {D.}~\bibnamefont {{Wineland}}},\
  }\href {\doibase 10.1103/RevModPhys.75.281} {\bibfield  {journal} {\bibinfo
  {journal} {Rev. Mod. Phys.}\ }\textbf {\bibinfo {volume} {75}},\ \bibinfo
  {pages} {281} (\bibinfo {year} {2003})}\BibitemShut {NoStop}%
\bibitem [{\citenamefont {{Turchette}}\ \emph {et~al.}(2000)\citenamefont
  {{Turchette}}, \citenamefont {{Kielpinski}}, \citenamefont {{King}},
  \citenamefont {{Leibfried}}, \citenamefont {{Meekhof}}, \citenamefont
  {{Myatt}}, \citenamefont {{Rowe}}, \citenamefont {{Sackett}}, \citenamefont
  {{Wood}}, \citenamefont {{Itano}}, \citenamefont {{Monroe}},\ and\
  \citenamefont {{Wineland}}}]{Turchette2000}%
  \BibitemOpen
  \bibfield  {author} {\bibinfo {author} {\bibfnamefont {Q.~A.}\ \bibnamefont
  {{Turchette}}}, \bibinfo {author} {\bibfnamefont {D.}~\bibnamefont
  {{Kielpinski}}}, \bibinfo {author} {\bibfnamefont {B.~E.}\ \bibnamefont
  {{King}}}, \bibinfo {author} {\bibfnamefont {D.}~\bibnamefont {{Leibfried}}},
  \bibinfo {author} {\bibfnamefont {D.~M.}\ \bibnamefont {{Meekhof}}}, \bibinfo
  {author} {\bibfnamefont {C.~J.}\ \bibnamefont {{Myatt}}}, \bibinfo {author}
  {\bibfnamefont {M.~A.}\ \bibnamefont {{Rowe}}}, \bibinfo {author}
  {\bibfnamefont {C.~A.}\ \bibnamefont {{Sackett}}}, \bibinfo {author}
  {\bibfnamefont {C.~S.}\ \bibnamefont {{Wood}}}, \bibinfo {author}
  {\bibfnamefont {W.~M.}\ \bibnamefont {{Itano}}}, \bibinfo {author}
  {\bibfnamefont {C.}~\bibnamefont {{Monroe}}}, \ and\ \bibinfo {author}
  {\bibfnamefont {D.~J.}\ \bibnamefont {{Wineland}}},\ }\href {\doibase
  10.1103/PhysRevA.61.063418} {\bibfield  {journal} {\bibinfo  {journal} {Phys.
  Rev. A}\ }\textbf {\bibinfo {volume} {61}},\ \bibinfo {eid} {063418}
  (\bibinfo {year} {2000})}\BibitemShut {NoStop}%
\bibitem [{\citenamefont {{Bruzewicz}}\ \emph {et~al.}(2015)\citenamefont
  {{Bruzewicz}}, \citenamefont {{Sage}},\ and\ \citenamefont
  {{Chiaverini}}}]{Bruzewicz2015}%
  \BibitemOpen
  \bibfield  {author} {\bibinfo {author} {\bibfnamefont {C.~D.}\ \bibnamefont
  {{Bruzewicz}}}, \bibinfo {author} {\bibfnamefont {J.~M.}\ \bibnamefont
  {{Sage}}}, \ and\ \bibinfo {author} {\bibfnamefont {J.}~\bibnamefont
  {{Chiaverini}}},\ }\href {\doibase 10.1103/PhysRevA.91.041402} {\bibfield
  {journal} {\bibinfo  {journal} {Phys. Rev. A}\ }\textbf {\bibinfo {volume}
  {91}},\ \bibinfo {eid} {041402} (\bibinfo {year} {2015})}\BibitemShut
  {NoStop}%
\bibitem [{\citenamefont {{Boldin}}\ \emph {et~al.}(2018)\citenamefont
  {{Boldin}}, \citenamefont {{Kraft}},\ and\ \citenamefont
  {{Wunderlich}}}]{Boldin2017}%
  \BibitemOpen
  \bibfield  {author} {\bibinfo {author} {\bibfnamefont {I.~A.}\ \bibnamefont
  {{Boldin}}}, \bibinfo {author} {\bibfnamefont {A.}~\bibnamefont {{Kraft}}}, \
  and\ \bibinfo {author} {\bibfnamefont {C.}~\bibnamefont {{Wunderlich}}},\
  }\href {\doibase 10.1103/PhysRevLett.120.023201} {\bibfield  {journal}
  {\bibinfo  {journal} {Phys. Rev. Lett.}\ }\textbf {\bibinfo {volume} {120}},\
  \bibinfo {eid} {023201} (\bibinfo {year} {2018})}\BibitemShut {NoStop}%
\bibitem [{\citenamefont {{Brownnutt}}\ \emph {et~al.}(2015)\citenamefont
  {{Brownnutt}}, \citenamefont {{Kumph}}, \citenamefont {{Rabl}},\ and\
  \citenamefont {{Blatt}}}]{Brownnutt2014}%
  \BibitemOpen
  \bibfield  {author} {\bibinfo {author} {\bibfnamefont {M.}~\bibnamefont
  {{Brownnutt}}}, \bibinfo {author} {\bibfnamefont {M.}~\bibnamefont
  {{Kumph}}}, \bibinfo {author} {\bibfnamefont {P.}~\bibnamefont {{Rabl}}}, \
  and\ \bibinfo {author} {\bibfnamefont {R.}~\bibnamefont {{Blatt}}},\ }\href
  {\doibase 10.1103/RevModPhys.87.1419} {\bibfield  {journal} {\bibinfo
  {journal} {Rev. Mod. Phys.}\ }\textbf {\bibinfo {volume} {87}},\ \bibinfo
  {pages} {1419} (\bibinfo {year} {2015})}\BibitemShut {NoStop}%
\bibitem [{\citenamefont {{Kielpinski}}\ \emph {et~al.}(2000)\citenamefont
  {{Kielpinski}}, \citenamefont {{King}}, \citenamefont {{Myatt}},
  \citenamefont {{Sackett}}, \citenamefont {{Turchette}}, \citenamefont
  {{Itano}}, \citenamefont {{Monroe}}, \citenamefont {{Wineland}},\ and\
  \citenamefont {{Zurek}}}]{Kielpinski2000}%
  \BibitemOpen
  \bibfield  {author} {\bibinfo {author} {\bibfnamefont {D.}~\bibnamefont
  {{Kielpinski}}}, \bibinfo {author} {\bibfnamefont {B.~E.}\ \bibnamefont
  {{King}}}, \bibinfo {author} {\bibfnamefont {C.~J.}\ \bibnamefont {{Myatt}}},
  \bibinfo {author} {\bibfnamefont {C.~A.}\ \bibnamefont {{Sackett}}}, \bibinfo
  {author} {\bibfnamefont {Q.~A.}\ \bibnamefont {{Turchette}}}, \bibinfo
  {author} {\bibfnamefont {W.~M.}\ \bibnamefont {{Itano}}}, \bibinfo {author}
  {\bibfnamefont {C.}~\bibnamefont {{Monroe}}}, \bibinfo {author}
  {\bibfnamefont {D.~J.}\ \bibnamefont {{Wineland}}}, \ and\ \bibinfo {author}
  {\bibfnamefont {W.~H.}\ \bibnamefont {{Zurek}}},\ }\href {\doibase
  10.1103/PhysRevA.61.032310} {\bibfield  {journal} {\bibinfo  {journal} {Phys.
  Rev. A}\ }\textbf {\bibinfo {volume} {61}},\ \bibinfo {pages} {032310}
  (\bibinfo {year} {2000})}\BibitemShut {NoStop}%
\bibitem [{\citenamefont {{Yudin}}\ and\ \citenamefont
  {{Taichenachev}}(2018)}]{Yudin2017}%
  \BibitemOpen
  \bibfield  {author} {\bibinfo {author} {\bibfnamefont {V.~I.}\ \bibnamefont
  {{Yudin}}}\ and\ \bibinfo {author} {\bibfnamefont {A.~V.}\ \bibnamefont
  {{Taichenachev}}},\ }\href {\doibase 10.1088/1612-202X/aa9aa5} {\bibfield
  {journal} {\bibinfo  {journal} {Laser Phys. Lett.}\ }\textbf {\bibinfo
  {volume} {15}},\ \bibinfo {pages} {035703} (\bibinfo {year}
  {2018})}\BibitemShut {NoStop}%
\bibitem [{\citenamefont {Major}\ \emph {et~al.}(2005)\citenamefont {Major},
  \citenamefont {Gheorge},\ and\ \citenamefont {Werth}}]{Major2005}%
  \BibitemOpen
  \bibfield  {author} {\bibinfo {author} {\bibfnamefont {F.~G.}\ \bibnamefont
  {Major}}, \bibinfo {author} {\bibfnamefont {V.~N.}\ \bibnamefont {Gheorge}},
  \ and\ \bibinfo {author} {\bibfnamefont {G.}~\bibnamefont {Werth}},\
  }\href@noop {} {\emph {\bibinfo {title} {{Charged Particle Traps}}}}\
  (\bibinfo  {publisher} {Springer},\ \bibinfo {year} {2005})\BibitemShut
  {NoStop}%
\bibitem [{\citenamefont {{Paul}}(1990)}]{Paul1990}%
  \BibitemOpen
  \bibfield  {author} {\bibinfo {author} {\bibfnamefont {W.}~\bibnamefont
  {{Paul}}},\ }\href {\doibase 10.1103/RevModPhys.62.531} {\bibfield  {journal}
  {\bibinfo  {journal} {Rev. Mod. Phys.}\ }\textbf {\bibinfo {volume} {62}},\
  \bibinfo {pages} {531} (\bibinfo {year} {1990})}\BibitemShut {NoStop}%
\bibitem [{\citenamefont {{Wineland}}\ \emph {et~al.}(1998)\citenamefont
  {{Wineland}}, \citenamefont {{Monroe}}, \citenamefont {{Itano}},
  \citenamefont {{Leibfried}}, \citenamefont {{King}},\ and\ \citenamefont
  {{Meekhof}}}]{Wineland1998}%
  \BibitemOpen
  \bibfield  {author} {\bibinfo {author} {\bibfnamefont {D.~J.}\ \bibnamefont
  {{Wineland}}}, \bibinfo {author} {\bibfnamefont {C.}~\bibnamefont
  {{Monroe}}}, \bibinfo {author} {\bibfnamefont {W.~M.}\ \bibnamefont
  {{Itano}}}, \bibinfo {author} {\bibfnamefont {D.}~\bibnamefont
  {{Leibfried}}}, \bibinfo {author} {\bibfnamefont {B.~E.}\ \bibnamefont
  {{King}}}, \ and\ \bibinfo {author} {\bibfnamefont {D.~M.}\ \bibnamefont
  {{Meekhof}}},\ }\href {\doibase 10.6028/jres.103.019} {\bibfield  {journal}
  {\bibinfo  {journal} {J. Res. Natl. Inst. Stand. Technol.}\ }\textbf
  {\bibinfo {volume} {103}},\ \bibinfo {pages} {259} (\bibinfo {year}
  {1998})}\BibitemShut {NoStop}%
\bibitem [{\citenamefont {{Ramsey}}\ and\ \citenamefont
  {{Silsbee}}(1951)}]{Ramsey1951}%
  \BibitemOpen
  \bibfield  {author} {\bibinfo {author} {\bibfnamefont {N.~F.}\ \bibnamefont
  {{Ramsey}}}\ and\ \bibinfo {author} {\bibfnamefont {H.~B.}\ \bibnamefont
  {{Silsbee}}},\ }\href {\doibase 10.1103/PhysRev.84.506} {\bibfield  {journal}
  {\bibinfo  {journal} {Phys. Rev.}\ }\textbf {\bibinfo {volume} {84}},\
  \bibinfo {pages} {506} (\bibinfo {year} {1951})}\BibitemShut {NoStop}%
\bibitem [{\citenamefont {{Peik}}\ \emph {et~al.}(2006)\citenamefont {{Peik}},
  \citenamefont {{Schneider}},\ and\ \citenamefont {{Tamm}}}]{Peik2006}%
  \BibitemOpen
  \bibfield  {author} {\bibinfo {author} {\bibfnamefont {E.}~\bibnamefont
  {{Peik}}}, \bibinfo {author} {\bibfnamefont {T.}~\bibnamefont {{Schneider}}},
  \ and\ \bibinfo {author} {\bibfnamefont {C.}~\bibnamefont {{Tamm}}},\ }\href
  {\doibase 10.1088/0953-4075/39/1/012} {\bibfield  {journal} {\bibinfo
  {journal} {J. Phys. B}\ }\textbf {\bibinfo {volume} {39}},\ \bibinfo {pages}
  {145} (\bibinfo {year} {2006})}\BibitemShut {NoStop}%
\bibitem [{\citenamefont {{Wimperis}}(1994)}]{Wimperis1994}%
  \BibitemOpen
  \bibfield  {author} {\bibinfo {author} {\bibfnamefont {S.}~\bibnamefont
  {{Wimperis}}},\ }\href {\doibase 10.1006/jmra.1994.1159} {\bibfield
  {journal} {\bibinfo  {journal} {J. Magn. Reson.}\ }\textbf {\bibinfo {volume}
  {109}},\ \bibinfo {pages} {221} (\bibinfo {year} {1994})}\BibitemShut
  {NoStop}%
\bibitem [{\citenamefont {{Porsev}}\ and\ \citenamefont
  {{Derevianko}}(2006)}]{Porsev2006}%
  \BibitemOpen
  \bibfield  {author} {\bibinfo {author} {\bibfnamefont {S.~G.}\ \bibnamefont
  {{Porsev}}}\ and\ \bibinfo {author} {\bibfnamefont {A.}~\bibnamefont
  {{Derevianko}}},\ }\href {\doibase 10.1103/PhysRevA.74.020502} {\bibfield
  {journal} {\bibinfo  {journal} {Phys. Rev. A}\ }\textbf {\bibinfo {volume}
  {74}},\ \bibinfo {eid} {020502} (\bibinfo {year} {2006})}\BibitemShut
  {NoStop}%
\bibitem [{\citenamefont {{Didier}}\ \emph {et~al.}(2018)\citenamefont
  {{Didier}}, \citenamefont {{Dole\v{z}al}}, \citenamefont {{Burgermeister}},
  \citenamefont {{Balling}},\ and\ \citenamefont
  {{Mehlst\"aubler}}}]{Didier2018}%
  \BibitemOpen
  \bibfield  {author} {\bibinfo {author} {\bibfnamefont {A.}~\bibnamefont
  {{Didier}}}, \bibinfo {author} {\bibfnamefont {M.}~\bibnamefont
  {{Dole\v{z}al}}}, \bibinfo {author} {\bibfnamefont {T.}~\bibnamefont
  {{Burgermeister}}}, \bibinfo {author} {\bibfnamefont {P.}~\bibnamefont
  {{Balling}}}, \ and\ \bibinfo {author} {\bibfnamefont {T.~E.}\ \bibnamefont
  {{Mehlst\"aubler}}},\ }\href@noop {} {\bibfield  {journal} {\bibinfo
  {journal} {in preparation for Appl. Phys. B}\ } (\bibinfo {year}
  {2018})}\BibitemShut {NoStop}%
\bibitem [{\citenamefont {{Dole\v{z}al}}\ \emph {et~al.}(2015)\citenamefont
  {{Dole\v{z}al}}, \citenamefont {{Balling}}, \citenamefont {{Nisbet-Jones}},
  \citenamefont {{King}}, \citenamefont {{Jones}}, \citenamefont {{Klein}},
  \citenamefont {{Gill}}, \citenamefont {{Lindvall}}, \citenamefont {{Wallin}},
  \citenamefont {{Merimaa}}, \citenamefont {{Tamm}}, \citenamefont {{Sanner}},
  \citenamefont {{Huntemann}}, \citenamefont {{Scharnhorst}}, \citenamefont
  {{Leroux}}, \citenamefont {{Schmidt}}, \citenamefont {{Burgermeister}},
  \citenamefont {{Mehlst\"aubler}},\ and\ \citenamefont
  {{Peik}}}]{Dolezal2015}%
  \BibitemOpen
  \bibfield  {author} {\bibinfo {author} {\bibfnamefont {M.}~\bibnamefont
  {{Dole\v{z}al}}}, \bibinfo {author} {\bibfnamefont {P.}~\bibnamefont
  {{Balling}}}, \bibinfo {author} {\bibfnamefont {P.~B.~R.}\ \bibnamefont
  {{Nisbet-Jones}}}, \bibinfo {author} {\bibfnamefont {S.~A.}\ \bibnamefont
  {{King}}}, \bibinfo {author} {\bibfnamefont {J.~M.}\ \bibnamefont {{Jones}}},
  \bibinfo {author} {\bibfnamefont {H.~A.}\ \bibnamefont {{Klein}}}, \bibinfo
  {author} {\bibfnamefont {P.}~\bibnamefont {{Gill}}}, \bibinfo {author}
  {\bibfnamefont {T.}~\bibnamefont {{Lindvall}}}, \bibinfo {author}
  {\bibfnamefont {A.~E.}\ \bibnamefont {{Wallin}}}, \bibinfo {author}
  {\bibfnamefont {M.}~\bibnamefont {{Merimaa}}}, \bibinfo {author}
  {\bibfnamefont {C.}~\bibnamefont {{Tamm}}}, \bibinfo {author} {\bibfnamefont
  {C.}~\bibnamefont {{Sanner}}}, \bibinfo {author} {\bibfnamefont
  {N.}~\bibnamefont {{Huntemann}}}, \bibinfo {author} {\bibfnamefont
  {N.}~\bibnamefont {{Scharnhorst}}}, \bibinfo {author} {\bibfnamefont {I.~D.}\
  \bibnamefont {{Leroux}}}, \bibinfo {author} {\bibfnamefont {P.~O.}\
  \bibnamefont {{Schmidt}}}, \bibinfo {author} {\bibfnamefont {T.}~\bibnamefont
  {{Burgermeister}}}, \bibinfo {author} {\bibfnamefont {T.~E.}\ \bibnamefont
  {{Mehlst\"aubler}}}, \ and\ \bibinfo {author} {\bibfnamefont
  {E.}~\bibnamefont {{Peik}}},\ }\href {\doibase 10.1088/0026-1394/52/6/842}
  {\bibfield  {journal} {\bibinfo  {journal} {Metrologia}\ }\textbf {\bibinfo
  {volume} {52}},\ \bibinfo {pages} {842} (\bibinfo {year} {2015})}\BibitemShut
  {NoStop}%
\bibitem [{\citenamefont {{Ludlow}}\ and\ \citenamefont
  {{Ye}}(2015)}]{Ludlow2015}%
  \BibitemOpen
  \bibfield  {author} {\bibinfo {author} {\bibfnamefont {A.~D.}\ \bibnamefont
  {{Ludlow}}}\ and\ \bibinfo {author} {\bibfnamefont {J.}~\bibnamefont
  {{Ye}}},\ }\href {\doibase 10.1016/j.crhy.2015.03.008} {\bibfield  {journal}
  {\bibinfo  {journal} {C. R. Phys.}\ }\textbf {\bibinfo {volume} {16}},\
  \bibinfo {pages} {499} (\bibinfo {year} {2015})}\BibitemShut {NoStop}%
\bibitem [{\citenamefont {{Safronova}}\ \emph {et~al.}(2011)\citenamefont
  {{Safronova}}, \citenamefont {{Kozlov}},\ and\ \citenamefont
  {{Clark}}}]{Safronova2011}%
  \BibitemOpen
  \bibfield  {author} {\bibinfo {author} {\bibfnamefont {M.~S.}\ \bibnamefont
  {{Safronova}}}, \bibinfo {author} {\bibfnamefont {M.~G.}\ \bibnamefont
  {{Kozlov}}}, \ and\ \bibinfo {author} {\bibfnamefont {C.~W.}\ \bibnamefont
  {{Clark}}},\ }\href {\doibase 10.1103/PhysRevLett.107.143006} {\bibfield
  {journal} {\bibinfo  {journal} {Phys. Rev. Lett.}\ }\textbf {\bibinfo
  {volume} {107}},\ \bibinfo {pages} {143006} (\bibinfo {year}
  {2011})}\BibitemShut {NoStop}%
\bibitem [{\citenamefont {{Vutha}}\ \emph {et~al.}(2017)\citenamefont
  {{Vutha}}, \citenamefont {{Kirchner}},\ and\ \citenamefont
  {{Dub{\'e}}}}]{Vutha2017}%
  \BibitemOpen
  \bibfield  {author} {\bibinfo {author} {\bibfnamefont {A.~C.}\ \bibnamefont
  {{Vutha}}}, \bibinfo {author} {\bibfnamefont {T.}~\bibnamefont {{Kirchner}}},
  \ and\ \bibinfo {author} {\bibfnamefont {P.}~\bibnamefont {{Dub{\'e}}}},\
  }\href {\doibase 10.1103/PhysRevA.96.022704} {\bibfield  {journal} {\bibinfo
  {journal} {Phys. Rev. A}\ }\textbf {\bibinfo {volume} {96}},\ \bibinfo
  {pages} {022704} (\bibinfo {year} {2017})}\BibitemShut {NoStop}%
\bibitem [{\citenamefont {{Jau}}\ \emph {et~al.}(2015)\citenamefont {{Jau}},
  \citenamefont {{Hunker}},\ and\ \citenamefont {{Schwindt}}}]{Jau2015}%
  \BibitemOpen
  \bibfield  {author} {\bibinfo {author} {\bibfnamefont {Y.-Y.}\ \bibnamefont
  {{Jau}}}, \bibinfo {author} {\bibfnamefont {J.~D.}\ \bibnamefont {{Hunker}}},
  \ and\ \bibinfo {author} {\bibfnamefont {P.~D.~D.}\ \bibnamefont
  {{Schwindt}}},\ }\href {\doibase 10.1063/1.4935562} {\bibfield  {journal}
  {\bibinfo  {journal} {AIP Advances}\ }\textbf {\bibinfo {volume} {5}},\
  \bibinfo {eid} {117209} (\bibinfo {year} {2015})}\BibitemShut {NoStop}%
\bibitem [{\citenamefont {{Rosenband}}\ \emph {et~al.}(2008)\citenamefont
  {{Rosenband}}, \citenamefont {{Hume}}, \citenamefont {{Schmidt}},
  \citenamefont {{Chou}}, \citenamefont {{Brusch}}, \citenamefont {{Lorini}},
  \citenamefont {{Oskay}}, \citenamefont {{Drullinger}}, \citenamefont
  {{Fortier}}, \citenamefont {{Stalnaker}}, \citenamefont {{Diddams}},
  \citenamefont {{Swann}}, \citenamefont {{Newbury}}, \citenamefont {{Itano}},
  \citenamefont {{Wineland}},\ and\ \citenamefont
  {{Bergquist}}}]{Rosenband2008}%
  \BibitemOpen
  \bibfield  {author} {\bibinfo {author} {\bibfnamefont {T.}~\bibnamefont
  {{Rosenband}}}, \bibinfo {author} {\bibfnamefont {D.~B.}\ \bibnamefont
  {{Hume}}}, \bibinfo {author} {\bibfnamefont {P.~O.}\ \bibnamefont
  {{Schmidt}}}, \bibinfo {author} {\bibfnamefont {C.~W.}\ \bibnamefont
  {{Chou}}}, \bibinfo {author} {\bibfnamefont {A.}~\bibnamefont {{Brusch}}},
  \bibinfo {author} {\bibfnamefont {L.}~\bibnamefont {{Lorini}}}, \bibinfo
  {author} {\bibfnamefont {W.~H.}\ \bibnamefont {{Oskay}}}, \bibinfo {author}
  {\bibfnamefont {R.~E.}\ \bibnamefont {{Drullinger}}}, \bibinfo {author}
  {\bibfnamefont {T.~M.}\ \bibnamefont {{Fortier}}}, \bibinfo {author}
  {\bibfnamefont {J.~E.}\ \bibnamefont {{Stalnaker}}}, \bibinfo {author}
  {\bibfnamefont {S.~A.}\ \bibnamefont {{Diddams}}}, \bibinfo {author}
  {\bibfnamefont {W.~C.}\ \bibnamefont {{Swann}}}, \bibinfo {author}
  {\bibfnamefont {N.~R.}\ \bibnamefont {{Newbury}}}, \bibinfo {author}
  {\bibfnamefont {W.~M.}\ \bibnamefont {{Itano}}}, \bibinfo {author}
  {\bibfnamefont {D.~J.}\ \bibnamefont {{Wineland}}}, \ and\ \bibinfo {author}
  {\bibfnamefont {J.~C.}\ \bibnamefont {{Bergquist}}},\ }\href {\doibase
  10.1126/science.1154622} {\bibfield  {journal} {\bibinfo  {journal}
  {Science}\ }\textbf {\bibinfo {volume} {319}},\ \bibinfo {pages} {1808}
  (\bibinfo {year} {2008})}\BibitemShut {NoStop}%
\bibitem [{\citenamefont {{Ruster}}\ \emph {et~al.}(2016)\citenamefont
  {{Ruster}}, \citenamefont {{Schmiegelow}}, \citenamefont {{Kaufmann}},
  \citenamefont {{Warschburger}}, \citenamefont {{Schmidt-Kaler}},\ and\
  \citenamefont {{Poschinger}}}]{Ruster2016}%
  \BibitemOpen
  \bibfield  {author} {\bibinfo {author} {\bibfnamefont {T.}~\bibnamefont
  {{Ruster}}}, \bibinfo {author} {\bibfnamefont {C.~T.}\ \bibnamefont
  {{Schmiegelow}}}, \bibinfo {author} {\bibfnamefont {H.}~\bibnamefont
  {{Kaufmann}}}, \bibinfo {author} {\bibfnamefont {C.}~\bibnamefont
  {{Warschburger}}}, \bibinfo {author} {\bibfnamefont {F.}~\bibnamefont
  {{Schmidt-Kaler}}}, \ and\ \bibinfo {author} {\bibfnamefont {U.~G.}\
  \bibnamefont {{Poschinger}}},\ }\href {\doibase 10.1007/s00340-016-6527-4}
  {\bibfield  {journal} {\bibinfo  {journal} {Appl. Phys. B}\ }\textbf
  {\bibinfo {volume} {122}},\ \bibinfo {eid} {254} (\bibinfo {year}
  {2016})}\BibitemShut {NoStop}%
\bibitem [{\citenamefont {Ruster}\ \emph {et~al.}(2017)\citenamefont {Ruster},
  \citenamefont {Kaufmann}, \citenamefont {Luda}, \citenamefont {Kaushal},
  \citenamefont {Schmiegelow}, \citenamefont {Schmidt-Kaler},\ and\
  \citenamefont {Poschinger}}]{Ruster2017}%
  \BibitemOpen
  \bibfield  {author} {\bibinfo {author} {\bibfnamefont {T.}~\bibnamefont
  {Ruster}}, \bibinfo {author} {\bibfnamefont {H.}~\bibnamefont {Kaufmann}},
  \bibinfo {author} {\bibfnamefont {M.~A.}\ \bibnamefont {Luda}}, \bibinfo
  {author} {\bibfnamefont {V.}~\bibnamefont {Kaushal}}, \bibinfo {author}
  {\bibfnamefont {C.~T.}\ \bibnamefont {Schmiegelow}}, \bibinfo {author}
  {\bibfnamefont {F.}~\bibnamefont {Schmidt-Kaler}}, \ and\ \bibinfo {author}
  {\bibfnamefont {U.~G.}\ \bibnamefont {Poschinger}},\ }\href {\doibase
  10.1103/PhysRevX.7.031050} {\bibfield  {journal} {\bibinfo  {journal} {Phys.
  Rev. X}\ }\textbf {\bibinfo {volume} {7}},\ \bibinfo {pages} {031050}
  (\bibinfo {year} {2017})}\BibitemShut {NoStop}%
\bibitem [{Sup()}]{Supplemental}%
  \BibitemOpen
  \href@noop {} {}\bibinfo {note} {{See Supplemental Material at [URL will be
  inserted by publisher] for details on the experimental setup and
  evaluation}}\BibitemShut {NoStop}%
\bibitem [{\citenamefont {{Schulte}}\ \emph {et~al.}(2016)\citenamefont
  {{Schulte}}, \citenamefont {{L{\"o}rch}}, \citenamefont {{Leroux}},
  \citenamefont {{Schmidt}},\ and\ \citenamefont {{Hammerer}}}]{Schulte2015}%
  \BibitemOpen
  \bibfield  {author} {\bibinfo {author} {\bibfnamefont {M.}~\bibnamefont
  {{Schulte}}}, \bibinfo {author} {\bibfnamefont {N.}~\bibnamefont
  {{L{\"o}rch}}}, \bibinfo {author} {\bibfnamefont {I.~D.}\ \bibnamefont
  {{Leroux}}}, \bibinfo {author} {\bibfnamefont {P.~O.}\ \bibnamefont
  {{Schmidt}}}, \ and\ \bibinfo {author} {\bibfnamefont {K.}~\bibnamefont
  {{Hammerer}}},\ }\href {\doibase 10.1103/PhysRevLett.116.013002} {\bibfield
  {journal} {\bibinfo  {journal} {Phys. Rev. Lett.}\ }\textbf {\bibinfo
  {volume} {116}},\ \bibinfo {pages} {013002} (\bibinfo {year}
  {2016})}\BibitemShut {NoStop}%
\bibitem [{\citenamefont {{Roos}}\ \emph {et~al.}(2006)\citenamefont {{Roos}},
  \citenamefont {{Chwalla}}, \citenamefont {{Kim}}, \citenamefont {{Riebe}},\
  and\ \citenamefont {{Blatt}}}]{Roos2006}%
  \BibitemOpen
  \bibfield  {author} {\bibinfo {author} {\bibfnamefont {C.~F.}\ \bibnamefont
  {{Roos}}}, \bibinfo {author} {\bibfnamefont {M.}~\bibnamefont {{Chwalla}}},
  \bibinfo {author} {\bibfnamefont {K.}~\bibnamefont {{Kim}}}, \bibinfo
  {author} {\bibfnamefont {M.}~\bibnamefont {{Riebe}}}, \ and\ \bibinfo
  {author} {\bibfnamefont {R.}~\bibnamefont {{Blatt}}},\ }\href {\doibase
  10.1038/nature05101} {\bibfield  {journal} {\bibinfo  {journal} {Nature}\
  }\textbf {\bibinfo {volume} {443}},\ \bibinfo {pages} {316} (\bibinfo {year}
  {2006})}\BibitemShut {NoStop}%
\bibitem [{\citenamefont {{Lechner}}\ \emph {et~al.}(2016)\citenamefont
  {{Lechner}}, \citenamefont {{Maier}}, \citenamefont {{Hempel}}, \citenamefont
  {{Jurcevic}}, \citenamefont {{Lanyon}}, \citenamefont {{Monz}}, \citenamefont
  {{Brownnutt}}, \citenamefont {{Blatt}},\ and\ \citenamefont
  {{Roos}}}]{Lechner2016}%
  \BibitemOpen
  \bibfield  {author} {\bibinfo {author} {\bibfnamefont {R.}~\bibnamefont
  {{Lechner}}}, \bibinfo {author} {\bibfnamefont {C.}~\bibnamefont {{Maier}}},
  \bibinfo {author} {\bibfnamefont {C.}~\bibnamefont {{Hempel}}}, \bibinfo
  {author} {\bibfnamefont {P.}~\bibnamefont {{Jurcevic}}}, \bibinfo {author}
  {\bibfnamefont {B.~P.}\ \bibnamefont {{Lanyon}}}, \bibinfo {author}
  {\bibfnamefont {T.}~\bibnamefont {{Monz}}}, \bibinfo {author} {\bibfnamefont
  {M.}~\bibnamefont {{Brownnutt}}}, \bibinfo {author} {\bibfnamefont
  {R.}~\bibnamefont {{Blatt}}}, \ and\ \bibinfo {author} {\bibfnamefont
  {C.~F.}\ \bibnamefont {{Roos}}},\ }\href {\doibase
  10.1103/PhysRevA.93.053401} {\bibfield  {journal} {\bibinfo  {journal} {Phys.
  Rev. A}\ }\textbf {\bibinfo {volume} {93}},\ \bibinfo {eid} {053401}
  (\bibinfo {year} {2016})}\BibitemShut {NoStop}%
\bibitem [{\citenamefont {{Ejtemaee}}\ and\ \citenamefont
  {{Haljan}}(2017)}]{Ejtemaee2017}%
  \BibitemOpen
  \bibfield  {author} {\bibinfo {author} {\bibfnamefont {S.}~\bibnamefont
  {{Ejtemaee}}}\ and\ \bibinfo {author} {\bibfnamefont {P.~C.}\ \bibnamefont
  {{Haljan}}},\ }\href {\doibase 10.1103/PhysRevLett.119.043001} {\bibfield
  {journal} {\bibinfo  {journal} {Phys. Rev. Lett}\ }\textbf {\bibinfo {volume}
  {119}},\ \bibinfo {pages} {043001} (\bibinfo {year} {2017})}\BibitemShut
  {NoStop}%
\bibitem [{\citenamefont {{Scharnhorst}}\ \emph {et~al.}(2018)\citenamefont
  {{Scharnhorst}}, \citenamefont {{Cerrillo}}, \citenamefont {{Kramer}},
  \citenamefont {{Leroux}}, \citenamefont {{W{\"u}bbena}}, \citenamefont
  {{Retzker}},\ and\ \citenamefont {{Schmidt}}}]{Scharnhorst2017}%
  \BibitemOpen
  \bibfield  {author} {\bibinfo {author} {\bibfnamefont {N.}~\bibnamefont
  {{Scharnhorst}}}, \bibinfo {author} {\bibfnamefont {J.}~\bibnamefont
  {{Cerrillo}}}, \bibinfo {author} {\bibfnamefont {J.}~\bibnamefont
  {{Kramer}}}, \bibinfo {author} {\bibfnamefont {I.~D.}\ \bibnamefont
  {{Leroux}}}, \bibinfo {author} {\bibfnamefont {J.~B.}\ \bibnamefont
  {{W{\"u}bbena}}}, \bibinfo {author} {\bibfnamefont {A.}~\bibnamefont
  {{Retzker}}}, \ and\ \bibinfo {author} {\bibfnamefont {P.~O.}\ \bibnamefont
  {{Schmidt}}},\ }\href {\doibase 10.1103/PhysRevA.98.023424} {\bibfield
  {journal} {\bibinfo  {journal} {Phys. Rev. A}\ }\textbf {\bibinfo {volume}
  {98}},\ \bibinfo {eid} {023424} (\bibinfo {year} {2018})}\BibitemShut
  {NoStop}%
\bibitem [{\citenamefont {{Huntemann}}\ \emph {et~al.}(2012)\citenamefont
  {{Huntemann}}, \citenamefont {{Okhapkin}}, \citenamefont {{Lipphardt}},
  \citenamefont {{Weyers}}, \citenamefont {{Tamm}},\ and\ \citenamefont
  {{Peik}}}]{Huntemann2012}%
  \BibitemOpen
  \bibfield  {author} {\bibinfo {author} {\bibfnamefont {N.}~\bibnamefont
  {{Huntemann}}}, \bibinfo {author} {\bibfnamefont {M.}~\bibnamefont
  {{Okhapkin}}}, \bibinfo {author} {\bibfnamefont {B.}~\bibnamefont
  {{Lipphardt}}}, \bibinfo {author} {\bibfnamefont {S.}~\bibnamefont
  {{Weyers}}}, \bibinfo {author} {\bibfnamefont {C.}~\bibnamefont {{Tamm}}}, \
  and\ \bibinfo {author} {\bibfnamefont {E.}~\bibnamefont {{Peik}}},\ }\href
  {\doibase 10.1103/PhysRevLett.108.090801} {\bibfield  {journal} {\bibinfo
  {journal} {Phys. Rev. Lett.}\ }\textbf {\bibinfo {volume} {108}},\ \bibinfo
  {pages} {090801} (\bibinfo {year} {2012})}\BibitemShut {NoStop}%
\bibitem [{\citenamefont {{Hobson}}\ \emph {et~al.}(2016)\citenamefont
  {{Hobson}}, \citenamefont {{Bowden}}, \citenamefont {{King}}, \citenamefont
  {{Baird}}, \citenamefont {{Hill}},\ and\ \citenamefont
  {{Gill}}}]{Hobson2016}%
  \BibitemOpen
  \bibfield  {author} {\bibinfo {author} {\bibfnamefont {R.}~\bibnamefont
  {{Hobson}}}, \bibinfo {author} {\bibfnamefont {W.}~\bibnamefont {{Bowden}}},
  \bibinfo {author} {\bibfnamefont {S.~A.}\ \bibnamefont {{King}}}, \bibinfo
  {author} {\bibfnamefont {P.~E.~G.}\ \bibnamefont {{Baird}}}, \bibinfo
  {author} {\bibfnamefont {I.~R.}\ \bibnamefont {{Hill}}}, \ and\ \bibinfo
  {author} {\bibfnamefont {P.}~\bibnamefont {{Gill}}},\ }\href {\doibase
  10.1103/PhysRevA.93.010501} {\bibfield  {journal} {\bibinfo  {journal} {Phys.
  Rev. A}\ }\textbf {\bibinfo {volume} {93}},\ \bibinfo {eid} {010501}
  (\bibinfo {year} {2016})},\ \Eprint {http://arxiv.org/abs/1510.08144}
  {arXiv:1510.08144 [physics.atom-ph]} \BibitemShut {NoStop}%
\bibitem [{\citenamefont {{Sanner}}\ \emph {et~al.}(2018)\citenamefont
  {{Sanner}}, \citenamefont {{Huntemann}}, \citenamefont {{Lange}},
  \citenamefont {{Tamm}},\ and\ \citenamefont {{Peik}}}]{Sanner2018}%
  \BibitemOpen
  \bibfield  {author} {\bibinfo {author} {\bibfnamefont {C.}~\bibnamefont
  {{Sanner}}}, \bibinfo {author} {\bibfnamefont {N.}~\bibnamefont
  {{Huntemann}}}, \bibinfo {author} {\bibfnamefont {R.}~\bibnamefont
  {{Lange}}}, \bibinfo {author} {\bibfnamefont {C.}~\bibnamefont {{Tamm}}}, \
  and\ \bibinfo {author} {\bibfnamefont {E.}~\bibnamefont {{Peik}}},\ }\href
  {\doibase 10.1103/PhysRevLett.120.053602} {\bibfield  {journal} {\bibinfo
  {journal} {Physical Review Letters}\ }\textbf {\bibinfo {volume} {120}},\
  \bibinfo {eid} {053602} (\bibinfo {year} {2018})}\BibitemShut {NoStop}%
\bibitem [{\citenamefont {Yudin}\ \emph {et~al.}(2018)\citenamefont {Yudin},
  \citenamefont {Taichenachev}, \citenamefont {Basalaev}, \citenamefont
  {Zanon-Willette}, \citenamefont {{Mehlst\"aubler}}, \citenamefont {Pollock},
  \citenamefont {Shuker}, \citenamefont {Donley},\ and\ \citenamefont
  {Kitching}}]{Yudin2018}%
  \BibitemOpen
  \bibfield  {author} {\bibinfo {author} {\bibfnamefont {W.~I.}\ \bibnamefont
  {Yudin}}, \bibinfo {author} {\bibfnamefont {A.~V.}\ \bibnamefont
  {Taichenachev}}, \bibinfo {author} {\bibfnamefont {M.~Y.}\ \bibnamefont
  {Basalaev}}, \bibinfo {author} {\bibfnamefont {T.}~\bibnamefont
  {Zanon-Willette}}, \bibinfo {author} {\bibfnamefont {T.~E.}\ \bibnamefont
  {{Mehlst\"aubler}}}, \bibinfo {author} {\bibfnamefont {J.~W.}\ \bibnamefont
  {Pollock}}, \bibinfo {author} {\bibfnamefont {M.}~\bibnamefont {Shuker}},
  \bibinfo {author} {\bibfnamefont {E.~A.}\ \bibnamefont {Donley}}, \ and\
  \bibinfo {author} {\bibfnamefont {J.}~\bibnamefont {Kitching}},\ }\href@noop
  {} {\bibfield  {journal} {\bibinfo  {journal} {ArXiv e-prints}\ } (\bibinfo
  {year} {2018})},\ \Eprint {http://arxiv.org/abs/1807.10158}
  {arXiv:1807.10158} \BibitemShut {NoStop}%
\bibitem [{\citenamefont {{Josephson}}(1960)}]{Josephson1960}%
  \BibitemOpen
  \bibfield  {author} {\bibinfo {author} {\bibfnamefont {B.~D.}\ \bibnamefont
  {{Josephson}}},\ }\href {\doibase 10.1103/PhysRevLett.4.341} {\bibfield
  {journal} {\bibinfo  {journal} {Phys. Rev. Lett.}\ }\textbf {\bibinfo
  {volume} {4}},\ \bibinfo {pages} {341} (\bibinfo {year} {1960})}\BibitemShut
  {NoStop}%
\bibitem [{\citenamefont {{Dehn}}(1970)}]{Dehn1970}%
  \BibitemOpen
  \bibfield  {author} {\bibinfo {author} {\bibfnamefont {J.~T.}\ \bibnamefont
  {{Dehn}}},\ }\href {\doibase 10.1016/0375-9601(70)90299-9} {\bibfield
  {journal} {\bibinfo  {journal} {Phys. Lett. A}\ }\textbf {\bibinfo {volume}
  {32}},\ \bibinfo {pages} {239} (\bibinfo {year} {1970})}\BibitemShut
  {NoStop}%
\end{thebibliography}%

\begin{appendix}
\section{\label{massdefect}Equivalence of time dilation and mass defect shifts in RF ion traps}
The thermal time dilation shift can alternatively be derived as a consequence of the different effective masses between a ground-state ($m_g$) and an excited-state ion ($m_e$) \cite{Josephson1960,Dehn1970,Yudin2017},
\begin{equation}
\label{massdefect}
m_e=m_g+\frac{h\nu_0}{c^2}\;\textnormal{.}
\end{equation}
We consider the case of a single ion in a linear rf trap, as used for multi-ion operation in the results of the main text. In the axial direction, confinement is provided by a static potential $\Phi_\mathrm{dc}$. The corresponding trap frequency is inversely proportional to the square root of the ion mass:
\begin{equation}
e\Phi_\mathrm{dc}(z)=\frac{1}{2}m\omega_\mathrm{ax}^2z^2\quad\Rightarrow\quad\nu_\mathrm{ax,g/e}=\frac{C_\mathrm{ax}}{\sqrt{m_{g/e}}}\;\textnormal{.}
\end{equation}
For a mean phonon number $\bar{n}$, the mass defect therefore results in an additional energy difference of
\begin{equation}
\Delta E_\mathrm{mass,ax}=\left(\frac{1}{2}+\bar{n}\right)hC_\mathrm{ax}\left[\frac{1}{\sqrt{m_e}}-\frac{1}{\sqrt{m_g}}\right]\;\textnormal{.}
\end{equation}
Inserting (\ref{massdefect}) yields
\begin{align}
\Delta E_\mathrm{mass,ax}&=\left(\frac{1}{2}+\bar{n}\right)hC_\mathrm{ax}\left[\frac{1}{\sqrt{m_g+\frac{h\nu_0}{c^2}}}-\frac{1}{\sqrt{m_g}}\right]\\
&=\left(\frac{1}{2}+\bar{n}\right)h\nu_\mathrm{ax,g}\left[\frac{1}{\sqrt{1+\frac{h\nu_0}{m_gc^2}}}-1\right]\label{axialmassdefect}\;\textnormal{.}
\end{align}
Using $m_gc^2\gg h\nu_0$ to approximate $1/\sqrt{1+x}$ by $1-x/2$, we obtain for the fractional frequency shift
\begin{equation}
\frac{\Delta\nu_\mathrm{mass,ax}}{\nu_0}\approx-\frac{\frac{1}{2}\left(\frac{1}{2}+\bar{n}\right)h\nu_\mathrm{ax,g}}{m_gc^2}=-\frac{\left\langle E_\mathrm{kin,ax}\right\rangle}{m_gc^2}\label{axialmdshift}\;\textnormal{,}
\end{equation}
i.e.~the ratio of average kinetic energy to rest energy, which is equivalent to the time dilation shift.\\

Next, we adapt the derivation for a direction with ponderomotive confinement, i.e.~either of the radial directions. Neglecting static electric field contributions and using the $q$ parameter as defined in Eq.~\ref{mathieuq} of the main text, we get the mass dependence
\begin{equation}
\omega_\mathrm{rad,pm}=\frac{\Omega_\mathrm{rf}}{\sqrt{8}}\vert q\vert=\frac{e\kappa_\mathrm{rf}U_\mathrm{rf}}{\sqrt{2}d_0^2m\Omega_\mathrm{rf}}\quad\Rightarrow\quad\nu_\mathrm{rad,g/e}=\frac{C_\mathrm{rad}}{m_{g/e}}\;\textnormal{,}
\end{equation}
such that the equivalent to (\ref{axialmassdefect}) is
\begin{align}
\Delta E_\mathrm{mass,rad}&=\left(\frac{1}{2}+\bar{n}\right)h\nu_\mathrm{rad,g}\left[\frac{-1}{\frac{m_gc^2}{h\nu_0}+1}\right]\\
&\approx-\left(\frac{1}{2}+\bar{n}\right)h\nu_\mathrm{rad,g}\;\frac{h\nu_0}{m_gc^2}\;\textnormal{.}
\end{align}
The fractional frequency shift in this case is twice as high as that arising from axial motion (\ref{axialmdshift}),
\begin{equation}
\frac{\Delta\nu_\mathrm{mass,rad}}{\nu_0}\approx-\frac{\left(\frac{1}{2}+\bar{n}\right)h\nu_\mathrm{rad,g}}{m_gc^2}=-2\frac{\left\langle E_\mathrm{kin,rad}\right\rangle}{m_gc^2}\;\textnormal{,}
\end{equation}
reflecting the additional kinetic energy due to intrinsic micromotion for a single ion, which corresponds to the potential energy of the harmonic oscillator with ponderomotive confinement.\\

For an equal temperature $T$ in the axial and both radial directions, $\langle E_\mathrm{kin,ax/rad}\rangle=k_BT/2$ and we recover the familiar thermal time dilation shift expression of $\Delta\nu/\nu_0=-5/2\times k_BT/(m_gc^2)$.
\end{appendix}

\end{document}